\documentclass[12pt]{article}          
\usepackage{comment}
\usepackage{amsmath,amssymb,graphicx,slashed}
\usepackage{braket} 
\usepackage{multirow}
\usepackage{feynmp-auto}
\usepackage{graphicx}                           
\usepackage{caption}
\usepackage{subcaption}                                                                
\graphicspath{ {images/} }
\usepackage{float}
\usepackage[tableposition=below]{caption}
\usepackage{longtable}
\usepackage{color}			
\usepackage{verbatim}
\usepackage[
	colorlinks=true,
	citecolor=blue,
	linkcolor=blue,
	urlcolor=blue,
	hypertexnames=false]{hyperref}

\newcommand{\LL}{\mathcal{L}}
\newcommand{\VV}{\mathcal{V}}
\newcommand{\OO}{\mathcal{O}}
\newcommand{\arXiv}[2]{\href{http://arxiv.org/pdf/#1}{{\tt #2/#1}}}
\newcommand{\arXivold}[1]{\href{http://arxiv.org/pdf/#1}{{\tt #1}}}
\newcommand{\beq}{\begin{eqnarray}}
\newcommand{\eeq}{\end{eqnarray}}
\numberwithin{equation}{section} 
\begin{document}
\begin{titlepage}
\vspace{1cm}
\begin{center}

	{	
		\LARGE \bf 
		Quantum Critical Higgs:\\}
		\vspace{0.1cm}
		{ \LARGE \bf From AdS$_5$ to Colliders 
	}
	
\end{center}
	\vskip .3cm
	\renewcommand*{\thefootnote}{\fnsymbol{footnote}}
\vspace{0.9cm}
\begin{center}
		
		\bf
		Ali Shayegan Shirazi\footnote{\tt \scriptsize
		 \href{mailto:shayegan@ucdavis.edu}{shayegan@ucdavis.edu},
		 $^\dag$\href{mailto: terning@physics.ucdavis.edu}{terning@physics.ucdavis.edu},
		 }, and John Terning$^{\dag}$
\end{center}
	
	\renewcommand{\thefootnote}{\arabic{footnote}}
	\setcounter{footnote}{0}

\begin{center} 

	{\it Department of Physics, University of California, One Shields Ave., Davis, CA 95616}

\end{center}

\vspace{1cm}

\centerline{\large\bf Abstract}
\begin{quote}
We examine distinctive signatures of Quantum Critical Higgs models at  the LHC and future higher energy colliders.
In these models the Higgs boson is part of a conformal sector that is softly broken at a threshold scale, and generically the scaling dimension of the Higgs is larger than in the Standard Model. In particular we examine the $gg\to H \to ZZ$, $gg\to H\to \gamma\gamma$, and $gg\to Z\to HZ$ channels to see how the cross sections deviate from the Standard Model in the high invariant mass region. In order to perform the calculations we use 5D duals of Quantum Critical Higgs models using the AdS/CFT correspondence, with a soft wall to break the conformal symmetry. 
 \end{quote}

\end{titlepage}


\section{Introduction}

 The major unsolved problem of the Standard Model (SM) is the hierarchy problem. This problem could have been resolved by technicolor, supersymmetry, or compositeness, but technicolor has been ruled out, and there is an abundant lack of  evidence for the latter two possibilities at low energies.  Thus we may need to explore new paradigms in order to find a solution, and it may be helpful to approach this problem using a different perspective---using the language of condensed matter physics. A condensed matter system that exhibits a phase transition at zero temperature as another  parameter of the system (like pressure or doping concentration) is changed, undergoes a quantum phase transition since there are only quantum fluctuations rather than the usual thermal fluctuations  \cite{sachdev}. Typically in such phase transitions we need to tune the parameters of the system to be close to the critical point. The Higgs sector of SM is in fact very similar to a Landau-Ginsburg model. The hierarchy problem is tantamount to the fact that in order for the Higgs VEV to be small we need to be extremely close to the critical point, and hence must fine-tune the parameters of the model. If the theory is a good description of nature up to the Planck scale, a one part in $10^{34}$ change of the Higgs mass term will push the theory far from its critical point. At the critical point, the Higgs mass vanishes , the correlation length diverges, and the low-energy effective theory is invariant under scale/conformal transformations. Due to the fact that the SM is weakly coupled, such a critical point is well approximated by mean-field theory, where the quantum fluctuations are small.
 
  In Quantum Critical Higgs (QCH) \cite{Bellazzini:2015cgj} models one considers critical points that do not exhibit mean-field behavior. That is, the  critical point  is described by a strongly coupled CFT. In such models, the scaling dimension of the Higgs field is different from its SM value $\Delta_{SM}=1+\OO (\alpha)$.
This assumption, however, does not automatically solve the hierarchy problem, though it might alleviate it since the Higgs field is really a composite. Here we will construct such models and analyze their observational consequences but postpone addressing the hierarchy problem. 
 
  The range of strongly coupled field theories in which we can actually do calculations without a lattice simulation is extremely limited and boils down to those with a weakly coupled dual, e.g. those that have a Seiberg-type dual or exhibit an AdS/CFT correspondence. Here we will focus on  the latter possibility; we will analyze two explicit 5D models that are candidates for a realistic QCH model.
  
   QCH models have some common features with the unparticles of Georgi \cite{Georgi:2007ek,Fox:2007sy,Cacciapaglia:2007jq,Delgado:2007dx,Kikuchi:2007qd,Delgado:2008rq,Delgado:2008px,Lee:2008ph,Espinosa:1998xj,vanderBij:2007um}. The conformal sector can show up in experiments as ``stuff" that is not like particles, but more like a continuum of particles \cite{Stancato:2008mp,Stancato:2008mp2,Falkowski:2008yr,Englert:2012dq,Englert:2012dq2,Goncalves:2018pkt,Csaki:2018kxb,Lee:2018fxj,Megias:2019vdb}. Several processes were suggested in ref.~\cite{Bellazzini:2015cgj} as candidates to test for  such new physics. The overall effect is an enhancement of certain cross sections above the threshold, $\mu$, where the effect of the continuum turns on. We will analyze the size of this enhancement is  in the channels $gg\to ZZ$, $gg\to \gamma\gamma$, and $gg\to HZ$. In ref.~\cite{Bellazzini:2015cgj}, it was predicted that there is a large enhancement in $gg\to ZZ$. We will see that this in fact is not the case. Among the processes we will study, perhaps the $gg\to HZ$ is the most promising channel in order to observe the effects of a QCH.

 In section \ref{secEffL} we describe some general features of QCH models in terms of the 1PI effective action. In section \ref{secAdSCFT} we present an AdS$_5$ version of a QCH model, show that it reproduces the SM at low energies, and discuss how to calculate non-trivial form factors from 5D. In section   \ref{sec:minimaladsCH} we present a second AdS$_5$ model that is simple enough that we can analytically calculate form factors,  include gauge interactions and higher dimensional interactions.  In section \ref{sec:collider}, we analyze the experimental signatures of the simple model,  both at the LHC and at a future 100 TeV collider. Finally we present our conclusions.
 
 
  \section{The 1PI effective action}\label{secEffL}
The action in QCH models has three parts,
\begin{equation}
S=S_{CFT}+S_{mix}+S_{SM}~.
\end{equation}
We are assuming that at least the Higgs and the longitudinal components of the gauge bosons are part of $S_{CFT}$, the conformal sector. $S_{SM}$ is the action for the other SM fields, and $S_{mix}$ contains the interaction between the two sectors.  
 
The dynamics of $S_{CFT}$ is unknown. To be compatible with the electroweak observations, however, the conformal symmetry should be broken near the TeV scale or higher. The breaking is characterized by a scale, $\mu$, that we refer to as the threshold. It is assumed to be a soft breaking of the CFT, as opposed to a hard breaking  which leads to confinement and Kaluza-Klein-like (KK-like)  resonances, as in Randall-Sundrum (RS) models \cite{RS}. We imagine that the theory flows \emph{close} to an IR fixed point, e.g. a Banks-Zaks fixed point, at some high energy scale, and the renormalization group flow stays close to this fixed point over a large range of renormalization scales.\footnote{For attempts to relate  such fixed points to the hierarcy problem see refs.~ \cite{HoldomCFT,walking,Agashe:2004rs}.} At the fixed point,  the theory is scale invariant, and throughout this paper we will assume that this point is also conformally invariant. Near the scale $\mu$ the flow moves away from the fixed point and below this scale the low-energy degrees of freedom  are those of the SM.

Assuming that there is such an approximate  CFT, we can try to find the form of the 1PI effective action that one would obtain by integrating over the CFT degrees of freedom, excluding the lightest modes (i.e. the Higgs and longitudinal gauge bosons). In momentum space the effective action has the following form
 \begin{equation}\label{unhiggsaction}
 S_{1PI}=\frac{1}{2\mathcal{Z}_h}\int \frac{d^4p}{(2\pi)^4}h(p)\Sigma (p^2,\mu, m_h,\Delta,\dots) h(-p)\quad+\quad \dots,
\end{equation}
 where $\Delta$ is the scaling dimension of the Higgs and $m_h$ the Higgs pole mass. The quadratic term, $\Sigma$, can also depend on other parameters which are either specific to a particular model, or are suppressed relative to the scale $\mu$. There are also self couplings of the Higgs and various interactions which we have not explicitly shown here and similar terms for the longitudinal polarizations of the gauge bosons. These fields will generally have non trivial form-factors in their interaction terms, as we will discuss later.
 
An assumption that we need to make is that the Higgs can be approximated by a generalized free field \cite{Greenberg} so that the effective action is weakly coupled when written in terms of these generalized free fields. This is familiar from the KK resonances in an RS theory, but in that case the anomalous dimension are required to be small as well. In the next section we will describe how we obtain such an effective action using the AdS/CFT correspondence, where the weakly-coupled 5D theory is dual to a strongly coupled CFT, and the perturbative parameter is $1/N$, the number of fields. But let us first discuss some of the general features of the quadratic term for the Higgs.
At the effective action level, all of our knowledge about the underlying CFT is encoded in the quadratic term and various form-factor interactions. We will concentrate here on the Higgs quadratic term and comment on gauge quadratic terms and the form-factors later. Some of the features discussed for the Higgs will apply to them as well.
 
For a generalized free scalar field, the dynamics of the CFT is effectively encoded in the scaling dimension $\Delta \ge 1$, where the bound comes from imposing unitarity \cite{Mack,Jterning}. The limit $\Delta =1$ corresponds to the case where the scalar is a free field, hence, should give back the SM (neglecting, for the moment, the Higgs pole mass): $ \Sigma(\Delta=1)=p^2$.  On the other hand, if the conformal breaking scale, $\mu$, is large then we should again approach the SM. 
  
  If there were no breaking of the conformal symmetry, the theory becomes purely conformal and the two-point function is fixed to be \cite{DiFrancesco:1997nk}
\begin{equation}
\braket{\mathcal{O}(x)\mathcal{O}(0)}\propto\frac{1}{x^{2\Delta}},
\end{equation}
while in momentum space we have
\begin{equation}\label{cft2pointfunction}
\braket{\OO(p)\OO(-p)}=\frac{-i}{p^{2(2-\Delta)}}~.
\end{equation}
Hence the scale invariant limit is $\Sigma(\mu=0,\dots)=p^{2(2-\Delta)}$, which should also apply more generally for $p^2 \gg \mu^2$. This limit corresponds to Georgi's unparticles. The non-integer power means that there is a branch cut in the two-point function starting at $p^2=0$, which is a signal that the field produces a continuum of states.
 
 As a consequence of the soft conformal breaking, the two-point function will have the branch cut starting from $\mu$. Equivalently, in the language of the K\"{a}ll\'{e}n-Lehmann spectral density \cite{Kallen:Lehmann,Kallen:Lehmann2},  the propagator is given by
\begin{align}
\braket{h(p)h(-p)}  &=-i \Sigma^{-1}\\
&  =\int_{0}^{\infty}\rho(M^2,\mu,\Delta)\frac{i}{p^2-M^2+i\epsilon}dM^2,\label{twopoint-densityfun}
\end{align}
where $\rho$ is the spectral density function. From the above equation we can write
\begin{equation}
\rho(p^2)  =-\frac{1}{\pi}\text{Im}\Bigg[\frac{1}{\Sigma(p^2+i \epsilon)}\Bigg] \label{densityfunction}.
\end{equation}
The spectral density is non-zero only above $\mu$. Such branch cuts are common in field theories, and they appear whenever the field produces two or more particles. In the SM, the Higgs has a branch cut starting from $p^2=0$ due to decay to photons. However, we will neglect this and other SM contributions to the spectral density since they are suppressed (by SM couplings) compared to that of production of (approximate) CFT states.

 So far we have ignored the Higgs mass. Ideally, one can require that the a mass term is generated, as a consequence of the conformal breaking, such that $m_h < \mu$, so that there are no excessive non-SM contributions  to the Higgs width, which we already know is very small. In this case the hierarchy problem is solved. However, we do not know how to guarantee this in the models we will be investigating; instead, we will fix the Higgs pole mass by coupling the QCH to an elementary scalar sector\footnote{In the language of AdS/CFT, this is tantamount to adding a UV mass term for the bulk Higgs. An AdS scenario could solve the hierarchy problem if there is a bulk mechanism that automatically gives a KK zero mode without requiring an additional UV mass term.}. Regardless of how the Higgs pole is generated, the phenomenology of the QCH will be the same. Hence, at this point we will simply restrict ourselves to the allowed region of parameter space and postpone addressing the Hierarchy problem to the future.
 
  Finally, we should note that a CFT in general admits any scaling dimension $\Delta \ge 1$; however, for $\Delta \ge 2$, a low-energy effective action for a QCH does not exist since the high momentum modes minimize the kinetic terms in the action. Furthermore, as explained in the next section, in an AdS description of a QCH  with $\Delta >2$ the QCH field would merely play the role of a source rather than being a dynamical field. For these reasons we will only consider scaling dimensions in the range $1<\Delta <2$.

\section{QCH from AdS$_5$} \label{secAdSCFT}

 At the end of the last century Maldacena conjectured that $\mathcal{N}= 4$ $SU(N)$ SUSY Yang-Mills in four dimensions, a conformally invariant theory, and Type IIB superstring theory on the $AdS_5\times S^5$ background are dual in the low-energy limit \cite{Maldacena:1997re}. Soon after, the correspondence was extended to the statement that any 5D gravity theory on an $AdS$ background could be dual to some CFT in four dimensions, for example, the RS model would correspond to a broken CFT \cite{ArkaniHamed:2000ds, Rattazzi:2000hs}.   
 A precise version of the conjecture is given by \cite{Witten:1998qj,Gubser,zaffaroni}
 \begin{equation}\label{wittenprescription}
 \big< e^{-\int d^4x\phi_0\mathcal{O}}\big>_{CFT}=e^{-S_{\text{5D}}[\phi]},
 \end{equation}
where on the left hand side we have the generating functional of the CFT connected correlation functions of operators $\mathcal{O}$, and on the right hand side we have the partition function of the classical 5D gravity theory. The value of the field $\phi$ at the UV boundary of the AdS space is $\phi_0$, which is the source for $\mathcal{O}$. We can impose a boundary condition in 5D for fields with $\Delta <2$ which promotes the boundary value to a dynamical field with a boundary action \cite{Klebanov:1999tb}; in the dual CFT picture the source becomes a dynamical field. Hence we can embed the SM in 5D by keeping all the SM fields on the boundary except those that are assumed to be part of a CFT, i.e. the Higgs and the longitudinal gauge modes.

The conformally flat 5D metric is
\begin{equation}\label{backgroundmetric}
ds^2=e^{-2 A(z)} \big(\eta_{\mu\nu}dx^{\mu}dx^{\nu}-dz^2\big),
\end{equation}
where $z$ is the coordinate of the extra dimension,  $\eta^{\mu\nu}$ the Minkowski metric with signature $(+---)$, 
and for pure AdS$_5$ we have $A(z)=\log z/R$,
with $R$ the curvature radius.

 Diffeomorphism invariance in the fifth dimension, the $\hat z$ direction in our coordinates, corresponds to the scale symmetry of the dual theory. Breaking this symmetry results in the breaking of the CFT in the dual theory.  A soft breaking of the CFT can be implemented using the so called soft-walls in 5D. Examples are a ``dilaton" field with a $z$-dependent VEV \cite{Batell:2008me} or a $z$-dependent mass \cite{Cacciapaglia:2008ns}, or a mixture of the two as in \cite{Falkowski:2008yr}. We will follow \cite{Batell:2008me} as the benchmark model for a 5D dual of a QCH model. In the next section we will use the model in \cite{Cacciapaglia:2008ns} to find closed expression for the form-factors. Although the later model is not as complete as  of the former, it is a good approximation to a QCH.
 
 In the soft-wall approach \cite{Batell:2008me}, the 5D action for gauge and matter fields is of the form
 \begin{equation}
 S=\int d^5x\sqrt{g}e^{-\Phi}\LL.
 \end{equation}
The scalar field $\Phi$ is the ``dilaton", which plays a role analogous to the ``dilaton" background in string theory  \cite{Polchinski:1998rq}.  Given an addition scalar field with a  suitable potential, the authors of \cite{Batell:2008me} have shown that it is possible to solve the Einstein equations to obtain an AdS metric with  
 \begin{align}\label{eq:dilaton_profile}
 \Phi(z)=&(2\mu z)^{a}~.
\end{align}  
The mass of the $n$th KK mode, for large $n$ is
\begin{equation}
m_n^2 \propto n^{2-2/a}~.
\end{equation}
This model is of special interest since with $a=2$, the spectrum of KK modes shows a linear ``Regge" behavior, so the model displays features of a holographic dual for QCD \cite{Karch:2006pv}. Here, we are interested in the $a=1$ case, where the spectrum of the KK modes becomes continuous. In other words, the 4D dual fields have a continuum spectrum above the scale $\mu$. We can either use the ``Schr\"{o}dinger equation" trick (see appendix~\ref{ads/brokenCFT}) to confirm this, or check the effective 4D action.
 
 We will denote the usual custodial symmetry preserving \cite{Agashe:2003zs} bulk gauge group fields corresponding to, $SU(2)_L\times SU(2)_R\times U(1)_X$ , by $L^a_M$, $R^a_M$, and $X_M$, where $a=1,2,3$ and $M=0,1,2,3,4$.  The associated gauge couplings are $g_{5L}$, $g_{5R}$, $g_X$ . 
 The 5D bulk action for the model is
 \begin{align}
 S_{5D}=&\int d^4xdz\sqrt{g}e^{-\Phi}\Bigg[-\frac{1}{4g_{5L}^2}{F^{a}_{L,MN}}^2-\frac{1}{4g_{5R}^2}{F^{a}_{R,MN}}^2-\frac{1}{4g_X^2}{F_{X,MN}}^2\Big]\nonumber \\
 &+\int d^4xdz\sqrt{g}e^{-\Phi}\Bigg[+|D_MH|^2-m^2H^2+\mathcal{L}_{int}(H)\Bigg]\nonumber \\
 &+\int d^4xdz\sqrt{g}e^{-\Phi}\delta(z-\epsilon) \mathcal{L}_{\text{UV}}.
 \end{align}
The Higgs and the Electroweak gauge fields propagate in the bulk, while the rest of SM fields live on the UV brane at $z=\epsilon$
 \begin{equation}	
 \LL_{\text{UV}}\supset \LL_{\text{rest of SM}}~.
 \end{equation}

Through the AdS/CFT dictionary \cite{ArkaniHamed:2000ds}, the fields living on the UV brane are elementary, while those propagating in the bulk are part of conformal sector. 
Also the bulk gauge symmetries correspond to the global symmetries of the CFT, some of which are weakly gauged, depending on the UV boundary conditions of the 5D theory. The UV boundary conditions \cite{Agashe:2003zs} that enforce a weakly gauged subgroup $SU(2)_L\times  U(1)_Y$, where $U(1)_1 \subset SU(2)_R\times U(1)_X$, are
\begin{align}\label{boundaryconditionsgaugefields1}
R_{\alpha}^{1,2}=&0,\\
X_{\alpha}-R_{\alpha}^3=&0,\\
\partial_z L_{\alpha}^a = & 0,\\
\partial_z \Big(\frac{1}{g_X^2}X_{\alpha}-\frac{1}{g_{5R}^2}R_{\alpha}^3\Big)=&0 \label{boundaryconditionsgaugefields2}.
\end{align}
Furthermore, the fifth components of the gauge fields are chosen to satisfy Dirichlet boundary conditions on the UV brane, and as a result, they do not have light modes in the 4D theory. 

The Higgs is a bi-doublet under $SU(2)_R\times SU(2)_L$, and adding a UV boundary potential
\begin{equation}
\LL_{UV}\supset-M_{UV}^2H_0^2+\lambda /4 H_0^4,
\end{equation}
allows for existence of a z-dependent solution, 
\begin{equation}\label{vev}
\VV(z)=\mathcal{K}(p=0,z)\VV_0,
\end{equation}
 such that the $\VV_0$ minimizes the 4D effective potential. Expanding around this solution and substituting
\begin{equation}
H=\frac{1}{\sqrt{2}}\begin{pmatrix},
0\\ h+\VV
\end{pmatrix}+\Pi
\end{equation}
we find 
\begin{equation}\label{fivedimensionalaction}
S_{5D}\supset  \frac{1}{2}\int d^4xdz\sqrt{g}e^{-\Phi}\Big[\partial_Mh\partial^Mh-m^2h^2+(\VV(z)+h)^2\Big(L_M^a-R_M^a\Big)^2+\dots \Big],
\end{equation}
where the dots represent mixing of gauge fields with Goldstone bosons, $\Pi$, and other Higgs interactions.

  The bulk mass, $m$, is related to the scaling dimension of the Higgs in the dual theory through the usual AdS/CFT dictionary \cite{Gubser},
 \begin{align}\label{scalingdimension}
\Delta=&2\pm\nu,\\
\nu=&\sqrt{m^2R^2+4}~.
\end{align}
 The ambiguity in (\ref{scalingdimension}) for the range $0<\nu<1$ will be discussed shortly. 
 
 The equation of motion for the Higgs, after Fourier transforming in 4D, is
\begin{equation}\label{bulkhiggsequationofmotion}
h''(p,z)+\big(\Phi'(z)-3A'(z)\big)h'(p,z)+\big(p^2-e^{-2A(z)}m^2\big)h(p,z)=0~,
\end{equation} 
where $^\prime$ indicates differentiation with respect to $z$.

 We can write the bulk field in terms of the fields on the UV brane using the so-called bulk-to-boundary propagator 
\begin{equation}\label{bulkscalar}
h(p,z)=h_{\epsilon}(p)\mathcal{K}(p,z)~.
\end{equation}
The propagator satisfies the same equation of motion as the Higgs, with a regularity condition at $z \to \infty$ (IR) and $\mathcal{K}=1$ at $z=\epsilon$ (UV).   With the ``dilaton" profile given by \eqref{eq:dilaton_profile}, we find
 \begin{align}
 \mathcal{K}(p,z)=& e^{(\mu-\sqrt{\mu^2-p^2})(z-\epsilon)}\Big(\frac{z}{\epsilon}\Big)^{2+\nu}\frac{U(\frac{1}{2}+\frac{3\mu}{2\sqrt{\mu^2-p^2}}+\nu,1+2\nu,2\sqrt{\mu^2-p^2}z)}{U(\frac{1}{2}+\frac{3\mu}{2\sqrt{\mu^2-p^2}}+\nu,1+2\nu,2\sqrt{\mu^2-p^2}\epsilon)},
 \end{align}
where $U$ is the confluent hypergeometric function \cite{Abramowitz:1965}.

Substituting $h$ into the action \eqref{fivedimensionalaction}, integrating by parts over the fifth dimension, and applying the equation of motion we obtain the holographic 4D action
\begin{equation}\label{effectiveAction}
S_{4D}=\int \frac{d^4p}{(2\pi)^4}\frac{R^3}{\epsilon^3}h_{\epsilon}(p) \Big[\partial_z\mathcal{K}(p,z=\epsilon)-m_{UV}^2\epsilon^{-2\Delta}\Big]h_{\epsilon}(-p),
\end{equation}
where $ m_{UV}^2=M_{UV}^2-3\lambda v_0$.

 We can now apply \eqref{wittenprescription} to find the connected correlation functions. However, it is well known that there is an ambiguity (\ref{scalingdimension}) for bulk masses in the range $0<\sqrt{m^2R^2+4}<1$, where $1<\Delta<2$ or $2<\Delta<3$. To select the solution with smaller scaling dimension, we need to do an additional Legendre transformation on the generating function \cite{Klebanov:1999tb}. In this case the boundary value $\phi_0$ is no longer a source, but a 4D field \cite{Cacciapaglia:2008ns}. That is, the effective action we get through the correspondence \eqref{effectiveAction} is the  1PI effective action for the CFT field with $\Delta=2-\nu$.

\begin{figure}[ht!]
\begin{center}    
\includegraphics[width=0.8\textwidth]{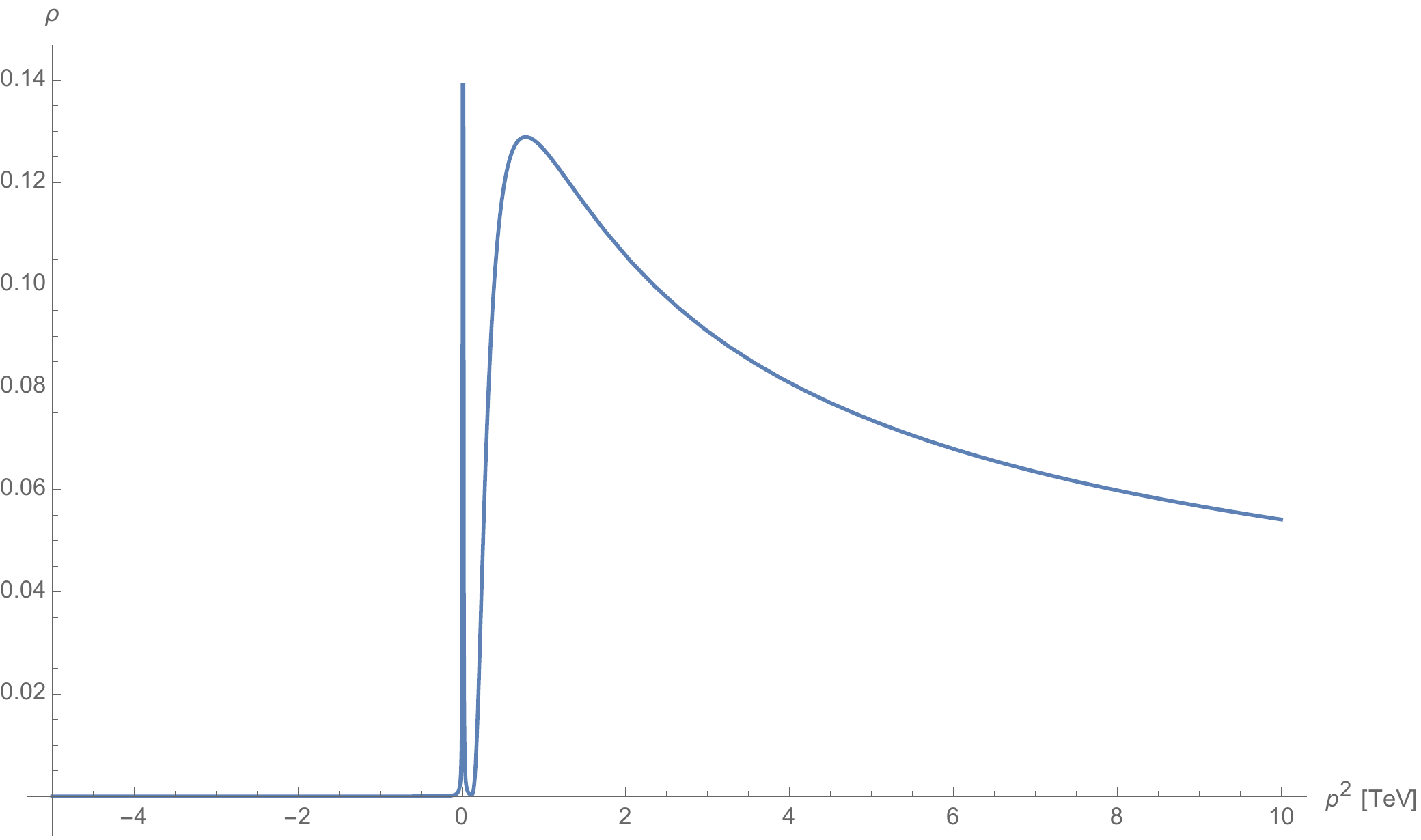}
    \caption{The Higgs spectral density function, for $\mu=300$ GeV and $\Delta=1.3$.}
        \label{densitygh}

\end{center}    
    \end{figure}

 After the rescaling
\begin{align}
 h_{\epsilon}^2=-&\frac{\epsilon^{4-2\nu}\Gamma(2\nu)}{2^{2\nu}R^3\Gamma(-2\nu)} h_{QCH}^2,
\end{align}
we can find the quadratic term 
 \begin{equation}\label{truedensityfunction}
 \Sigma(p^2)=-(\mu^2-p^2)^{\nu}\frac{\Gamma(\frac{1}{2}+\frac{3\mu}{2\sqrt{\mu^2-p^2}}+\nu)}{\Gamma(\frac{1}{2}+\frac{3\mu}{2\sqrt{\mu^2-p^2}}-\nu)}-m_{UV}^2.
\end{equation}  
The UV mass determines the Higgs pole mass through the relation
\begin{equation}
m_{UV}^2=-(\mu^2-m_{\text{h}}^2)^{\nu}\frac{\Gamma(\frac{1}{2}+\frac{3\mu}{2\sqrt{\mu^2-m_{h}^2}}+\nu)}{\Gamma(\frac{1}{2}+\frac{3\mu}{2\sqrt{\mu^2-m_{h}^2}}-\nu)}~.
\end{equation}
 The spectral density function,  \eqref{densityfunction}, for $\mu=300$ GeV and $\Delta=1.3$ is shown in figure~\ref{densitygh}.
  
     Turning to the gauge fields, we note that the 5D Higgs VEV \eqref{vev} breaks the $SU(2)_L\times SU(2)_R$ symmetry in the bulk and gives mass to one linear combination of the gauge fields: $L^a_M-R_M^a$. Hence we define the fields that diagonalize the mass matrix as
 \begin{align}
A_M^a&\equiv L^a_M-R_M^a,\\
V_M^a&\equiv L^a_M+R_M^a.
 \end{align}

     We need to add a gauge fixing term \cite{Randall:2001gb} to separate the equations of motion of the almost flat component of gauge fields, $A^a_{\mu}$, from the equations of motion of the fifth component, $A^a_5$. The gauge fixing term is
\begin{equation}
S_{5D-GF}=\int -\frac{1}{2g_5^2\xi}e^{-A(z)-\Phi}\Big(\partial_{M}A^{aM}-\xi e^{A(z)+\Phi(z)}\partial_z\Big(e^{-A(z)-\Phi(z)}A^a_5\Big)\Big)^2,
\end{equation}
with an identical term for $V^a$ as well. We will work in $\xi=1$ gauge.

     The 5D profile of the 4D Fourier transform of $A^a_{\mu}$ and $V^a_{\mu}$, are denoted by $a(p,z)$ and $v(p,z)$:
\begin{align}
A^a_{\alpha}=&A_{\alpha}^{a0}(p)a(p,z)\\
V^a_{\alpha}=&V_{\alpha}^{a0}(p)v(p,z)
\end{align}
 where $A^0$ and $V^0$ fix the UV boundary value of the fields. The profiles satisfy the equations of motion:
\begin{align}\label{eq:eomgauge}
\partial_z \Big(e^{-A(z)-\Phi(z)}\partial_z \Big)a(p,z)&-e^{-3A(z)-\Phi(z)}g_5^2\mathcal{V}(z)^2a(p,z)=-p^2e^{-A(z)-\Phi(z)}a(p,z)\nonumber \\
\partial_z \Big(e^{-A(z)-\Phi(z)}\partial_z \Big)v(p,z)&=-p^2e^{-A(z)-\Phi(z)}v(p,z)~.
\end{align}

The second equation can be solved. The solution that does not blow up in the IR is 
\begin{equation}
v(q,z)=e^{(\mu-\sqrt{\mu^2-q^2})(z-R)}\frac{z^2}{R^2}\frac{U\Big(\frac{\mu}{2\sqrt{\mu^2-q^2}}+3/2,3,2 z\sqrt{\mu^2-q^2}\Big)}{U\Big(\frac{\mu}{2\sqrt{\mu^2-q^2}}+3/2,3,2 z\sqrt{\mu^2-q^2}\Big)}~.
\end{equation}
The 4D effective action is 
\begin{equation}
S_{4D}=\int d p^4 \Big(\eta^{\alpha\beta}-p^{\alpha}p^{\beta}/p^2\Big)e^{-A(R)-\Phi(R)}V^0_{\alpha}(p)\partial_zv(p,z)|_{z=R}V^0_{\beta}(-p)~.
\end{equation}
In the limit $R\to 0$, we have
\begin{eqnarray}
\partial_zv(p,z)|_{z=R}&\approx& p^2 R\left[3/2-2\gamma -\log [2\sqrt{\mu^2-p^2}R] \right. \nonumber \\
&&\left.-\psi_0\Big(\frac{\mu}{2\sqrt{\mu^2-p^2}}+3/2\Big)-\frac{\mu^2-\mu\sqrt{\mu^2-p^2}}{p^2}\right], \label{quadratic:gauge}
\end{eqnarray}
where $\psi_0$ is the Poly-Gamma function \cite[Eq.~5.4.1]{Abramowitz:1965}.

\begin{figure}
\begin{center}
\includegraphics[width=0.8\textwidth]{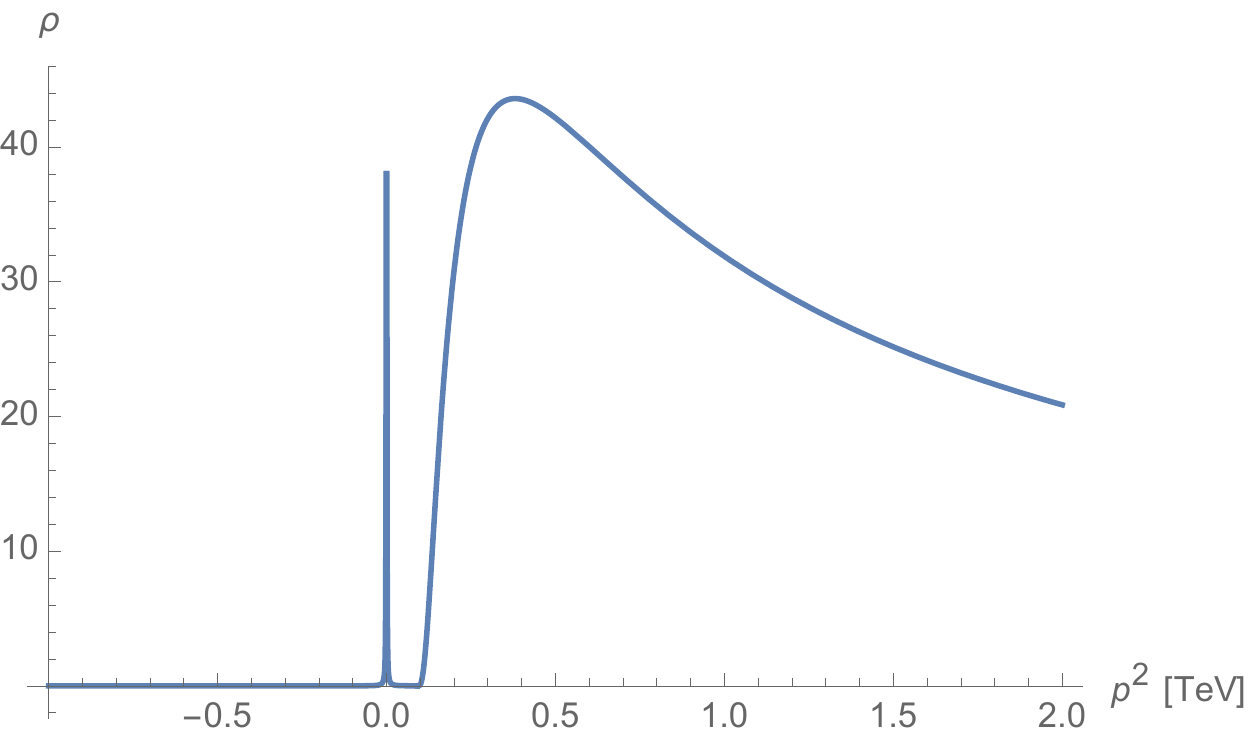}
\caption{The $V_\mu$ spectral density function, with $R=10^{-3}$, and $ \mu=0.3$ TeV.}
\label{photondensityfunction}
\end{center}
\end{figure}

In order to find SM fields, one needs to switch back to $L^a_{\mu}$ and $R^a_{\mu}$ basis and find the mass eigenstates and fix the parameters of the model. However, there is an easier way of making sure that we have the correct masses for the  $W$ and $Z$ gauge bosons in the spectrum. By applying the boundary conditions (\ref{boundaryconditionsgaugefields1}-\ref{boundaryconditionsgaugefields2})  we find three equations:
\begin{align}
v'(p,z_0)&=0~,\\
v(p,z_0)&a'(p,z_0)+a(p,z_0)v'(p,z_0)=0~,\label{eq:wspectrum}\\
v(p,z_0)&a'(p,z_0)+a(p,z_0)v'(p,z_0)+2\frac{g_X^2}{g_5^2}v(p,z_0)a'(p,z_0)=0~.\label{eq:zspectrum}
\end{align}
 The first equation corresponds to the photon, and has a solution with $p=\sqrt{p^2}=0$. The eigenvalue of the second equation, corresponding to the $W$, is smaller than that of the third equation. We can use the second equation at $p=m_W$ to fix $g_5^2\VV_0^2$. Knowing $g_5^2\VV_0^2$, we can then find $g_X^2/g_5^2$ for which the third equation is satisfied for $p=M_Z$. At the final step, one can use the gauge couplings  to fermions, which live on the UV brane, to find $g_5$.
 The gauge boson spectral density \eqref{densityfunction} is given by the imaginary part of the inverse of the quadratic term \eqref{quadratic:gauge}. A plot of the spectral density is shown in figure~\ref{photondensityfunction} with $R=10^{-3}$, and $ \mu=0.3$ TeV.

In summary, given the masses of the $W$ and $Z$ bosons, the pole mass of Higgs boson, and $\alpha=e^2/4\pi$, we can fix the 5D parameters $g_5,g_X,M_{UV}$, and $\lambda$. As a reminder, the parameters $m_{UV}$ and $\VV_0$ are given in terms of $M_{UV}$ and $\lambda$. The two parameters $\epsilon$ and $R$, which control the UV cut-off, are taken to be small, of order $10^{-2} \text{TeV}^{-1}$. The remaining parameters, $m$, related to the scaling dimension of the Higgs field in the 4D theory, and $\mu$,  are free.

\begin{figure}[h]
    \vspace{1cm}
   \begin{eqnarray*}
   \parbox{40mm}{\begin{fmffile}{hzzeffv3}\begin{fmfgraph*}(85,45)
\fmfleft{i1}
\fmfright{o1,o2}
\fmf{dbl_dashes}{i1,v1}
\fmf{photon}{v1,o1}
\fmf{photon}{v1,o2}
\fmfblob{0.15w}{v1}
\end{fmfgraph*}\end{fmffile}}\quad = &\quad
\parbox{40mm}{ \begin{fmffile}{wittenHZZv1}
\begin{fmfgraph}(60,60)
	\fmfsurroundn{v}{6}
	\fmfcyclen{dbl_plain,right=0.25}{v}{6}
\fmffreeze
	\fmf{dashes}{v4,o}
          \fmf{photon}{o,v2}
          \fmf{photon}{o,v6}
          \fmfdot{v2}
          \fmfdot{v4}
          \fmfdot{v6}
\end{fmfgraph}
        \end{fmffile}} + \quad\quad \parbox{40mm}{\begin{fmffile}{wittenhphizzbulkv1}
\begin{fmfgraph}(60,60)
	\fmfsurroundn{v}{12}
	\fmfcyclen{dbl_plain,right=0.127}{v}{12}
\fmffreeze
	\fmf{dashes}{v7,o1}
          \fmf{photon}{o2,v2}
          \fmf{photon}{o2,v12}
          \fmf{dots}{o1,o2}
          \fmfv{decor.shape=cross,decor.size=10}{o1}
          \fmfdot{v7}
          \fmfdot{v2}
          \fmfdot{v12}
\end{fmfgraph}
\end{fmffile}}
\end{eqnarray*}
\vspace{.4cm}
\caption{The effective interaction in the 4D theory (left) from propagation of the fields in the bulk and mixing with Goldstones (denoted by dotted line) in the bulk (right). We have denoted the 4D Higgs with double dashed lines}
 \label{effectiveHZZfromthebulk}
\end{figure}
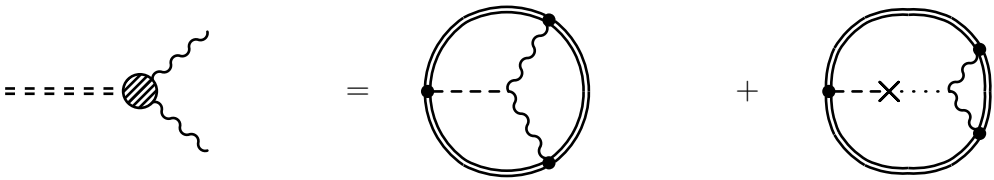

  The 4D interaction terms can be found by starting with the corresponding interactions in the bulk and propagating to the UV boundary. For the Higgs-$ZZ$ interaction, for example,  we can start with the third term in \eqref{fivedimensionalaction}  and find 
\begin{eqnarray*}\label{hzzfromthebulk}
S_{4D}=(4\pi)^4\delta(p+q_1+q_2)v_{\epsilon}h_{\epsilon}(p)Z_{\alpha}(q_1)Z_{\beta}(q_2)f^{\alpha\beta}(p,q_1,q_2),
\end{eqnarray*}
where 
\begin{equation}\label{eq:hzzintegral}
f^{\alpha\beta}(p,q_1,q_2)=\eta^{\alpha\beta}\int_{\epsilon}^{Z_{IR}} dz e^{-3A(z)-\Phi}\mathcal{K}(0,z)\mathcal{K}(p^2,z)\frac{a(q_1,z)a(q_2,z)}{a(q_1,z_0)a(q_2,z_0)}~.
\end{equation}
 In diagrammatic form, these interactions are shown using Witten diagrams in figure~\ref{effectiveHZZfromthebulk}; the middle diagram denotes the bulk integration in \eqref{hzzfromthebulk}. At tree level, other diagrams that contribute  are due to the mixing of the gauge fields with Goldstones and are proportional to $q_{1,2}^{\alpha}q_{1,2}^{\beta}$. One can find these interactions by using the Bulk-to-Bulk propagator for the Goldstones. For simplified models these interactions can be found using the Mandelstam method of gauging a non-local actions \cite{Mandelstam,Mandelstam2,Mandelstam3,Cacciapaglia:2007jq}, which we turn to in the next section.
 
\begin{figure}[th]
 \centering
 \includegraphics[width=.8\textwidth]{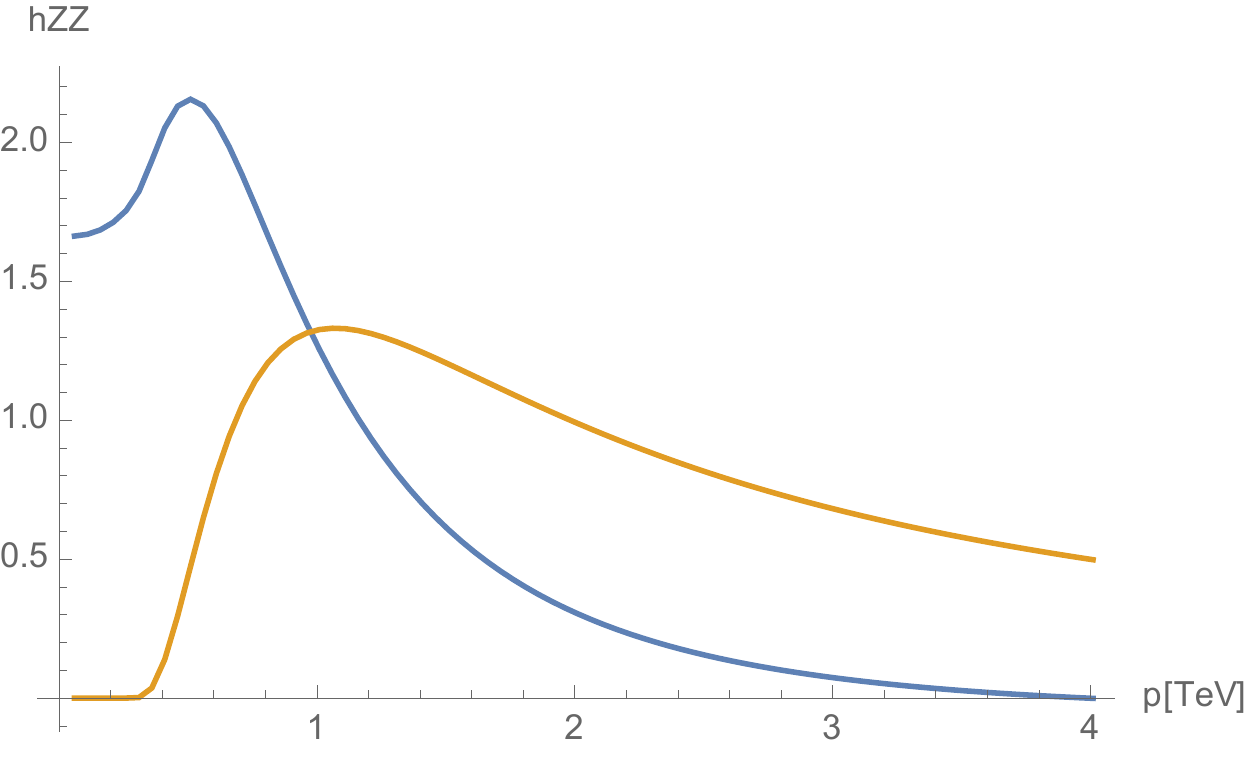}
 \caption{Real (blue) and imaginary (Orange) part of the hZZ form factor, with $\mu~=~0.2$ TeV, $\nu~=~0.3$. The imaginary part is only non-zero for $p>\mu$.}
 \label{gr:hzzGH}
  \end{figure}

  We have  numerically calculated the integral \eqref{eq:hzzintegral} for $R=\epsilon=10^{-3} \text{TeV}^{-1}$, $\Delta=1.7$, and $\mu=0.2$ TeV, as shown in figure~\ref{gr:hzzGH}.
 
\section{Minimal Coupling of AdS-QCH}\label{sec:minimaladsCH}

In this section we follow the model in ref.~\cite{Cacciapaglia:2008ns}. Due to the fact that it is possible to compute the form factors analytically, we will use this model later on in our collider analysis. The 5D bulk action is
 \begin{align}
 S_{5D}=&\int d^4xdz\sqrt{g}e^{-\Phi}\Bigg[-\frac{1}{4g_{5L}^2}{F^{a}_{L,MN}}^2-\frac{1}{4g_{5R}^2}{F^{a}_{R,MN}}^2-\frac{1}{4g_X^2}{F_{X,MN}}^2\Big]\nonumber \\
 &+\int d^4xdz\sqrt{g}\Bigg[+|D_MH|^2-\left[m^2+\mu \phi(z)\right]H^2+\mathcal{L}_{int}(H)\Bigg]\nonumber\\
 &+\int d^4x \mathcal{L}_{\text{UV}},
 \end{align}
where $\Phi$ is the same ``dilaton" as the previous section and has been added to give the gauge fields their thresholds.

\begin{figure}[h]
 \centering
 \includegraphics[width=0.8\textwidth]{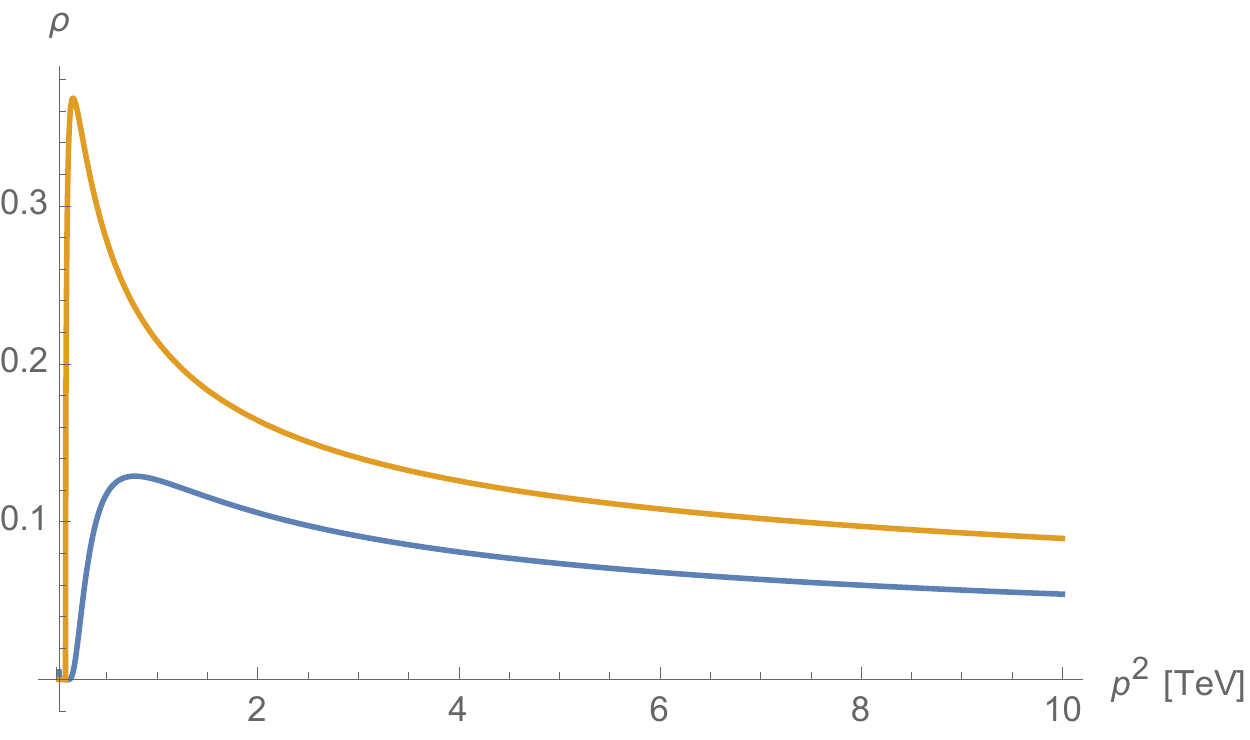}
 \caption{The spectral density function for the Minimal model as given by \eqref{miminaldensityfunction} (Orange, upper) and the spectral density functions for the first model as given by \eqref{truedensityfunction} (Blue, lower). $\mu=0.3$ TeV, $\Delta=2-\nu=1.7$.}
 \label{gr:density_compare}
 \end{figure}

In the Higgs sector there is no ``dilaton", instead the Higgs is coupled to a scalar field, $\phi$, with a z-dependent expectation value that gives the Higgs its threshold. We have
\begin{align}
\phi(z)=&\mu \Big(\frac{z}{R}\Big)^2.\label{phi}
\end{align}
 The threshold again \cite{Cacciapaglia:2008ns} starts at $\mu$ . The Bulk-Boundary propagator after requiring regularity in the IR is
 \begin{align}
\mathcal{K}(p,z)=&\Big(\frac{z}{\epsilon}\Big)^2\frac{K_{\nu}(\sqrt{\mu^2-p^2}z)}{K_{\nu}(\sqrt{\mu^2-p^2}\epsilon)},
 \end{align}
	where $K$ is the modified Bessel function of the second kind.
 After the following the same procedure discussed in the previous section, we find
 \begin{align}
 h_{\epsilon}^2= & \frac{\epsilon^{4-2\nu}2^{2\nu}\Gamma(\nu)}{R^3\Gamma(1-\nu)}h^2, \label{eq:redefinitionTR}\\
 \Sigma(p^2)=&-(\mu^2-p^2)^{\nu}+(\mu^2-m_h^2)^\nu\label{miminaldensityfunction}.
 \end{align}
  This spectral density function is compared to that of the previous model in figure~\ref{gr:density_compare}.
 
 The growing VEV of $\phi$, acts like a mass for the the Higgs field, and one might suspect that the back reaction of this VEV on the metric might form some sort of a singularity at some finite value of $z_s$. This can be disastrous since we know that singularities break the continuum and reproduce KK modes with mass splitting of order $\pi/z_s$. In appendix~\ref{app:backreaction} we show that even when this happens, the location of the singularity can be such that the mass splitting is less than the Higgs width, and so effectively the spectrum remains continuous.

The spectrum of the gauge fields are controlled by the ``dilaton" with the same profile and equation of motion as in the previous model. However contrary to the previous model the threshold for the gauge fields does not have to be the same as the Higgs threshold. Hence, the threshold of the gauge fields can be much larger, so we only have to deal with the zero modes. The mixing with the Goldstones however changes the longitudinal part of the gauge fields,  depending on the gauge choice. In unitary gauge the propagator is given by 
\begin{equation}
\frac{1}{p^2-m^2-i m  \Gamma_{\text{SM}}}\Bigg(-\eta^{\alpha\beta}+\frac{p^{\alpha}p^{\beta}}{p^2}\Big(1+\frac{p^2}{m^2\mu^2}\frac{(2-\Delta)(p^2-m^2)}{1-\mu^{2\Delta-4}(\mu^2-p^2)^{2-\Delta}}\Big)\Bigg)~.
\end{equation}

The integral in equation \eqref{eq:hzzintegral} with a flat gauge profile is easily done. After field redefinitions the form factor reduces to 
  \begin{align}
f^{\alpha\beta}(p,q_1,q_2) \approx  \frac{2\eta^{\alpha\beta}}{\mathcal{Z}_h} \frac{\mu^{2\nu}-(\mu^2-p^2)^{\nu}}{p^2}~.
\label{flatformfactor}
 \end{align}
 This is in fact just what we expect from minimal coupling and gauge invariance {\em a}  {\em la} Mandelstam \cite{Mandelstam,Mandelstam2,Mandelstam3}.  Due to mixing through the Higgs VEV in Eq.~\eqref{eq:eomgauge}, the profile of the zero modes are not exactly flat, however, the deviation starts approximately where the warp factor suppresses the integrand, hence the deviation from (\ref{flatformfactor}) is small.

As an aside, let us examine some higher dimension operators in the bulk that contribute to the Higgs-two gauge coupling in the effective theory. These turn out to be suppressed, so we will not need them in the collider analysis. 
One such dimension 6 operator is the coupling of two Higgs fields with two Field strength tensors,
\begin{align}
\LL^{(6)}_{5D}&= g_6 H^2 F_{MN}F_{ST}g^{MS}g^{NT}\nonumber\\
&=2\, g_6 e^{4A(z)} H^2 \Big(\partial_{\alpha}A_{\beta}\partial^{\alpha}A^{\beta}-\partial_{\beta}A_{\alpha}\partial^{\alpha}A^{\beta}-\partial_{\alpha}A_{5}\partial^{\alpha}A_{5}\nonumber \\ 
&\quad\quad\quad\quad\quad\quad\quad +\partial_{z}A_{\alpha}\partial_{z}A^{\alpha}+2\partial_{\beta}A_{5}\partial_{z}A^{\beta}\Big)~.
\end{align}
The third and the last term do not involve the light approximate zero-modes due to the boundary condition on $A_5$. In the approximation that the gauge profiles are flat, the fourth term vanishes as well. Setting one Higgs to its VEV we find an effective 4D H-AA coupling in momentum space given by
\begin{align}
\LL^{(6)}_{4D}&=2v_{\epsilon}h_{\epsilon}(p)A_{\alpha}(q_1)A_{\beta}(q_2)(q_1\cdot q_2\eta^{\alpha\beta}-q_1^{\alpha}q_2^{\beta})g_6 \nonumber \\
&\quad \times \int_{\epsilon}^{Z_{IR}} dz e^{-A(z)}\mathcal{K}(0,z)\mathcal{K}(p^2,z)
\end{align}

The integral can be done with the help of Mathematica. After the field redefinition \eqref{eq:redefinitionTR} we find
\begin{align}
\LL^{(6)}_{4D}=&2\,v_{0}h(p)A_{\alpha}(q_1)A_{\beta}(q_2)(q_1\cdot q_2\eta^{\alpha\beta}-q_1^{\beta}q_2^{\alpha})\times \nonumber\\
&4\frac{g_6}{R^2}\frac{2\mu^2\big(\mu^{2\nu}-(\mu^2-p^2)^{\nu}\big)-p^2\big((\mu^2-p^2)^{\nu}(\nu-1)+\mu^{2\nu}(\nu+1)\big)}{p^6}
\end{align}
Note that as $p^2\to0$, the numerator falls as $p^6$, and cancel the divergence of the denominator. 

\section{QCH In collider physics} \label{sec:collider}

  One of the consequences of embedding the Higgs sector of the SM in a conformal theory is the appearance of non-trivial momentum dependence in form factors at Higgs vertices \cite{Bellazzini:2015cgj}, in contrast to the usual sub-leading contribution coming from loop corrections in the SM. Such deviations are a signature of non-mean-field theories. A general discussion of the properties of these form factors has been given in \cite{Bellazzini:2015cgj}. We have explored a few of these form factors using the AdS/CFT correspondence: below we will look for signals of QCH in a few channels that are being studied at the LHC and will be promising in future colliders.
  
 We have used {\tt Madgraph5\_aMC} (MG) \cite{Alwall:2014hca} to generate the amplitudes. The details of the implementation of the propagators and form factors can be found in appendix~\ref{app:qchmadgraph}. We will use the propagator of section \ref{sec:minimaladsCH}  and denote it with a double-dashed line Feynman diagram as shown in figure~\ref{Higgsprop}.  
 
   \begin{figure}[h!]
      \centering
   \begin{eqnarray*}
\parbox{20mm}{\begin{fmffile}{hprop}
        \begin{fmfgraph*}(100, 55)
         \fmfleft{l1,l2,l3}
          \fmfright{r1,r2,r3}
          \fmf{dbl_dashes}{l2,r2}
        \end{fmfgraph*}\end{fmffile}
    }
    \qquad\qquad = & \frac{if}{-(\mu^2-p^2)^{\nu}+(\mu^2-m^2)^{\nu}-i m f \Gamma_{\text{SM}}}
            \end{eqnarray*}
            \caption{The QCH propagator used in the simulations.}
                        \label{Higgsprop}
    \end{figure}

In order for the QCH propagator to have unit residue at its pole mass, $m$, we set the normalization constant to be
 \begin{equation}
f=\frac{\nu}{(\mu^2-m^2)^{1-\nu}}~.
 \end{equation}
 Close to the pole mass the QCH propagator mimics that of the SM Higgs.

   In the model of section \ref{secAdSCFT} the gauge bosons had the same threshold as the Higgs, while in the model of section \ref{sec:minimaladsCH} the thresholds are independent. To keep matters simple, we will examine the second model and set the threshold of the gauge bosons to be much larger than the available energy, i.e. we will work with SM gauge bosons. However because of the Higgs mechanism, the Higgs threshold shows up in the propagators of the longitudinal gauge bosons.  These can be computed using either AdS/CFT or by applying gauge invariance with minimal coupling as in \cite{Stancato:2008mp,Stancato:2008mp2,Stancato:2011vaa}. In unitary gauge we have \cite{Bellazzini:2015cgj}
   
   \begin{figure}[ht!]
      \centering
   \begin{eqnarray*}
\parbox{20mm}{\begin{fmffile}{zprop}
        \begin{fmfgraph*}(100, 55)
         \fmfleft{l1,l2,l3}
          \fmfright{r1,r2,r3}
          \fmf{dbl_wiggly}{l2,r2}
        \end{fmfgraph*}\end{fmffile}
    }
    \qquad\qquad = & \frac{i}{p^2-m^2-i m  \Gamma_{\text{SM}}}\Bigg(-\eta^{\alpha\beta}+\frac{p^{\alpha}p^{\beta}}{p^2}\Big(1+\frac{p^2}{m^2\mu^2}\frac{\nu(p^2-m^2)}{1-(1-p^2/\mu^2)^{\nu}}\Big)\Bigg)
            \end{eqnarray*}

    \end{figure}

 One can also find the form factors for minimal gauge interactions by imposing gauge invariance on the non-local action \eqref{unhiggsaction} using the Mandelstam method \cite{Mandelstam,Mandelstam2,Mandelstam3}. This has been worked out in detail in \cite{Stancato:2011vaa} for a general kinetic term for the Higgs. The $Z$ and $W$ bosons cubic coupling to the Higgs is given by
\begin{eqnarray}
 \parbox{39mm}{\begin{fmffile}{zzhiggs}
        \begin{fmfgraph*}(100, 55)
         \fmfleft{l1,l2,l3}
          \fmfright{r1,r2,r3}
	\fmflabel{$\alpha$}{r1}
	\fmflabel{$\beta$}{r3}
          \fmf{dbl_wiggly,label=$q_1$}{o,r1}
          \fmf{dbl_wiggly,label=$q_2$}{o,r3}
          \fmf{dbl_dashes,label=$p$}{l2,o}
	 \fmfblob{0.15w}{o}
        \end{fmfgraph*}\end{fmffile}}
\quad  =-if^{\alpha\beta}(p,q_q,q_2),
\end{eqnarray}
where
\begin{align}
f^{\alpha\beta}(p,q_q,q_2)=\frac{g^2\mathcal{V}}{4c_w^2\mathcal{Z}_h}\Big[&2\eta^{\alpha\beta}A(0,q_1+q_2)\nonumber\\
&+(2q_2+q_1)^{\alpha}q_2^{\beta}B(0,q_2,q_1+q_2)\nonumber\\
&+q_1^{\alpha}(2q_1+q_2)^{\beta}B(q_1+q_2,q_1,0)\Big]\label{HzzvertixfromMaldestam}
\end{align}
and
\begin{align}
A(p,q)=&\frac{K(p)-K(q)}{p^2-q^2}\label{stancatoDeffinitionOfA}\\
B(q,k,p)=&\frac{K(q)}{(q^2-p^2)(q^2-k^2)}-\frac{K(k)}{(q^2-k^2)(k^2-p^2)}+\frac{K(p)}{(q^2-p^2)(k^2-p^2)}\\
K(p)=&(\mu^2 -p^2)^\nu~.
\end{align}          
The fermion couplings of the Higgs and the gauge bosons are taken to be the same as in the SM; for a discussion of  $\Delta> 1.5$ and fermion couplings see ref.~\cite{Englert:2012dq,Englert:2012dq2}. 
    
   Whenever the QCH or the longitudinal components of the gauge bosons are restricted to be nearly on-shell by experimental cuts, we can use the usual SM propagators, as in figure~\ref{SMprops}.    
      \begin{figure}[h!]
      \centering
   \begin{eqnarray*}
   \parbox{20mm}{\begin{fmffile}{hpropsm}
        \begin{fmfgraph*}(100, 55)
         \fmfleft{l1,l2,l3}
          \fmfright{r1,r2,r3}
          \fmf{dashes}{l2,r2}
        \end{fmfgraph*}\end{fmffile}
    }
    \qquad\qquad = & \frac{i}{p^2-m^2-i m \Gamma_{\text{SM}}}\\
\parbox{20mm}{\begin{fmffile}{zpropsm}
        \begin{fmfgraph*}(100, 55)
         \fmfleft{l1,l2,l3}
          \fmfright{r1,r2,r3}
          \fmf{wiggly}{l2,r2}
        \end{fmfgraph*}\end{fmffile}
    }
    \qquad\qquad = & \frac{i\big(-\eta^{\alpha\beta}+p^{\alpha}p^{\beta}/p^2\big)}{p^2-m_V^2-i m_V  \Gamma_{\text{SM}}}
            \end{eqnarray*}
  \caption{SM propagators for the Higgs and gauge bosons.}       
            \label{SMprops}
    \end{figure}

 \subsection{$gg\to ZZ\to \ell\ell\ell\ell$}
   In this channel, the Higgs decays to two $Z$ bosons which subsequently decay to four leptons. This is one of the main channels for studying the Higgs at the LHC and is sometimes called the Golden Channel since it provides such a clean signal. Studies in the off-shell region are an interesting way to search for new particles and for bounding the Higgs width \cite{Caola:2013yja,Melnikov:2015laa,Bellazzini:2015cgj,Aaboud:2018puo,Goncalves:2018pkt}.  In the off-shell region with $\sqrt{s}>2M_Z$, at the lowest order, we need to take into account two diagrams, $gg\to H\to ZZ$ and $gg\to ZZ$ through a fermion box (i.e. continuum production of $ZZ$) as depicted in figure~\ref{fdiag:ggzz}. We have generated these diagrams in MG, with the two $Z$ bosons on-shell and with a minimum cut on their transverse momentum $p_T(Z)>1 $GeV (for avoiding $1/p_T$ singularities \cite{Campbell:2013una}).
  \begin{figure}[ht!]
      \centering
   \begin{eqnarray*}
\parbox{20mm}{\begin{fmffile}{ggzzbox1}
        \begin{fmfgraph*}(100, 55)
         \fmftop{t1,t4}
          \fmfbottom{b1,b4}
          \fmf{curly}{t1,t2}
          \fmf{photon}{t3,t4}
          \fmf{curly}{b1,b2}
          \fmf{photon}{b3,b4}
          \fmf{fermion}{t3,t2}
          \fmf{fermion,tension=0.3}{t2,b2}
         \fmf{fermion}{b2,b3}
          \fmf{fermion,tension=0.3}{b3,t3}
        \end{fmfgraph*}\end{fmffile}
    }
    \qquad\qquad +\quad\text{crossings}\quad+\qquad 
    \parbox{20mm}{\begin{fmffile}{ggzzhiggs1}
        \begin{fmfgraph*}(100, 55)
         \fmftop{t1,t4}
          \fmfbottom{b1,b4}
          \fmf{curly}{t1,t2}
          \fmf{curly}{b1,b2}
          \fmf{fermion,tension=0.3}{t2,b2}
          \fmf{fermion}{b2,m1}
          \fmf{fermion}{m1,t2}
          \fmf{dbl_dashes}{m1,m2}
          \fmf{photon}{m2,b4}
          \fmf{photon}{m2,t4}
	 \fmfblob{0.15w}{m2}
        \end{fmfgraph*}\end{fmffile}}
        \end{eqnarray*}
    \caption{The $gg\to ZZ$ amplitude.}
    \label{fdiag:ggzz}
    \end{figure}
 \\ 
 
\begin{figure}[ht!]
 \centering
\includegraphics[width=.6 \textwidth]{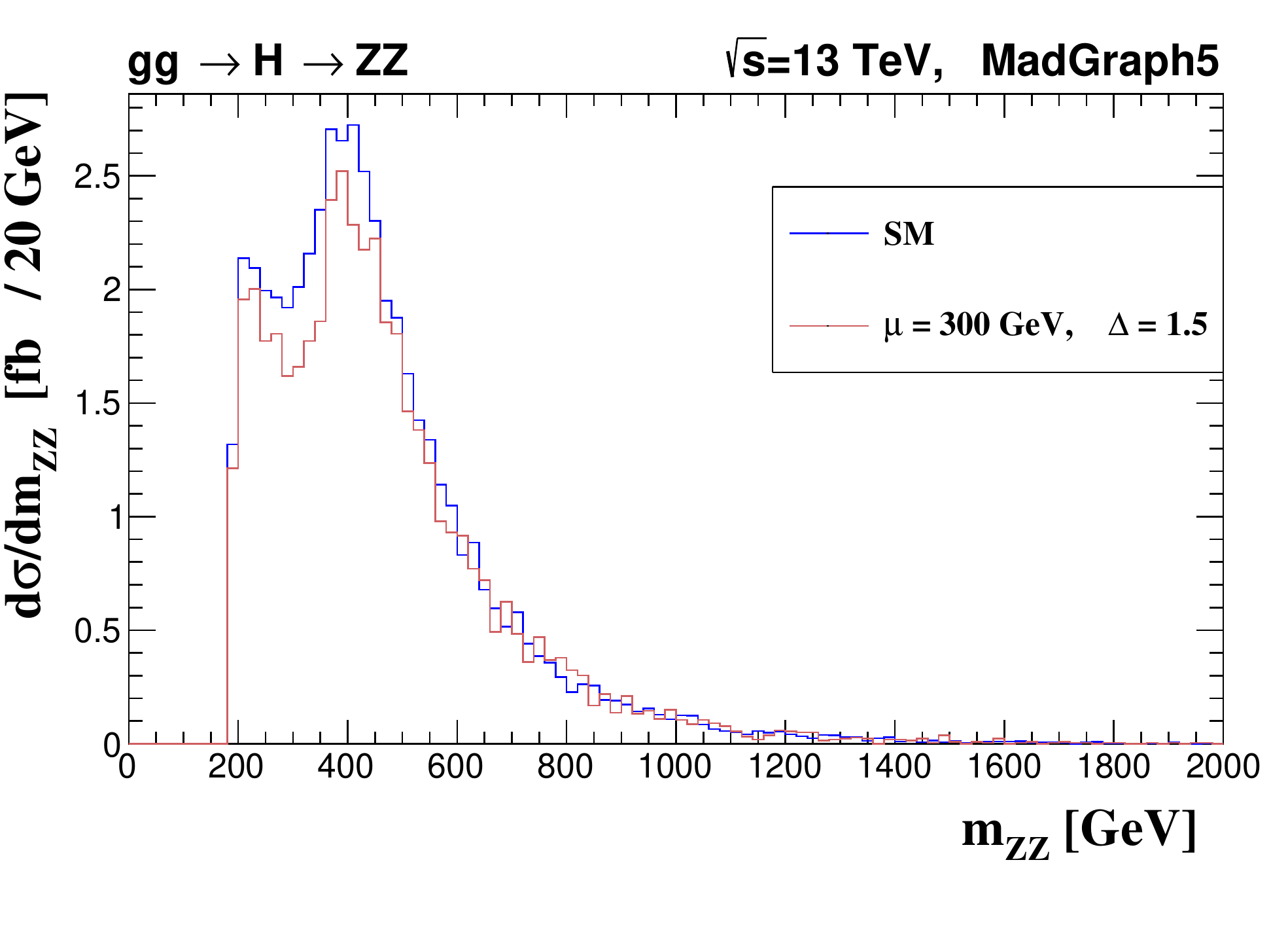}
\caption{$gg\to H \to ZZ$, with the $Z$ bosons on-shell.}
\label{fig:ggHzz}
\end{figure}

 For large invariant mass squared, $p^2\gg \mu^2$, the Higgs propagator falls as $1/p^{2\nu}$ and the $HZZ$  form factor falls as $1/p^{2-2\nu}$, so the QCH meditated amplitude falls just like SM as shown in figure~\ref{fig:ggHzz}. This is a consequence of gauge invariance.\footnote{As shown in the reference \cite{Englert:2019zmt}, when we add dimension 6 operators that modify the Higgs propagator, the new physics between the Higgs propagator and the gauge vertex cancels.}  

\begin{figure}
\centering
\includegraphics[width=0.7 \textwidth]{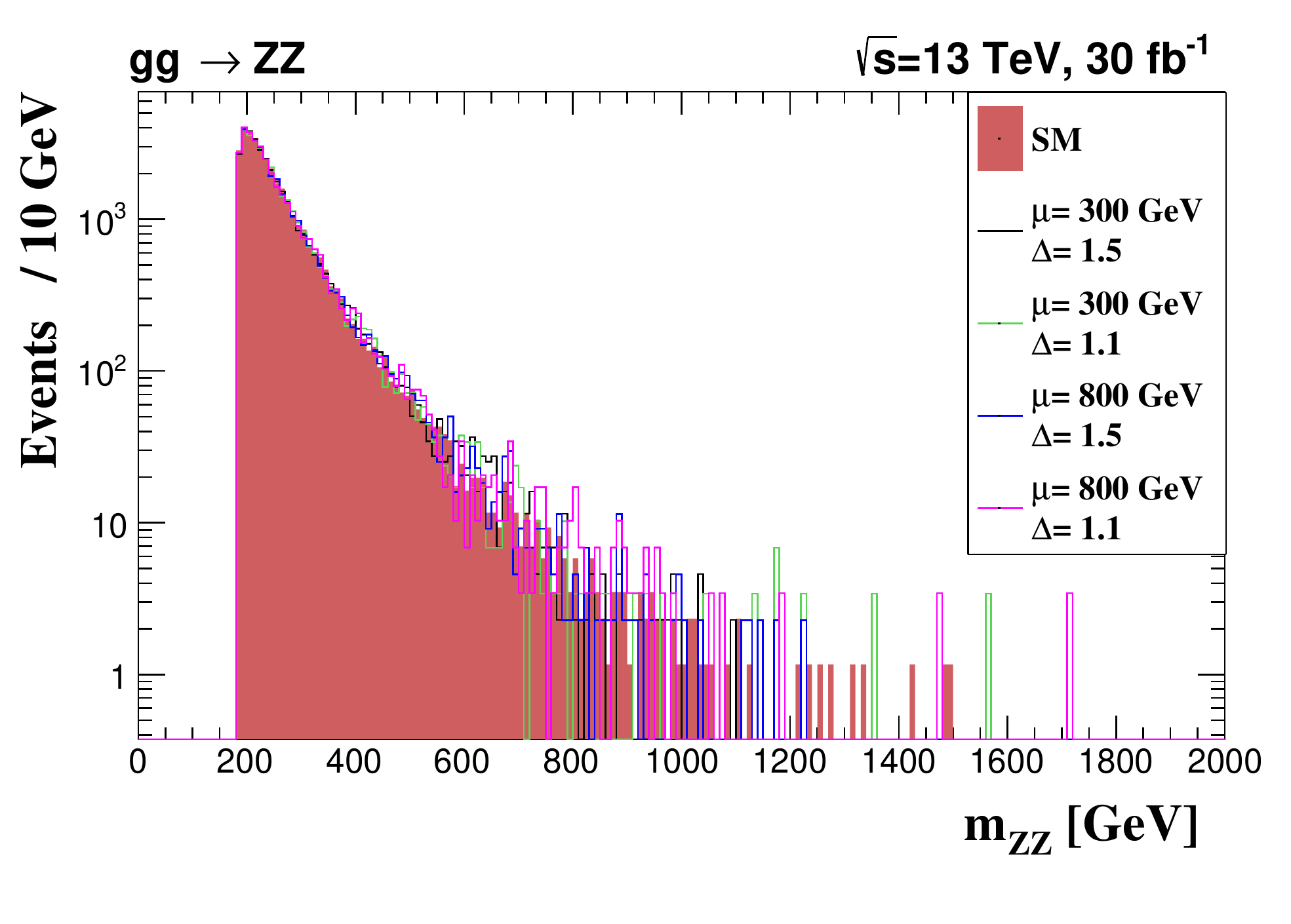}
  \caption{With the inclusion of $i\epsilon$ in the form factor (explained in the text), the strong deviation from SM above the threshold, $\mu$, vanishes. The fluctuation observed in the graph are due to using small number of events in MG. The $Z$'s are on-shell and with a cut on the $Z$ transverse momentum: $p_T>1$GeV.  (Color online) }
  \label{fig:ggzzMG}
\end{figure}

 Figure~\ref{fig:ggzzMG} shows the distribution of the events with a luminosity of 30 $ fb^{-1}$ using the interference of Higgs and continuum $Z$ production amplitudes. So the signal in this channel is quite modest and only significant at very high invariant masses with $m_{ZZ}\sim\mu$.
 We should note that this is in contradiction with the result of previous work \cite{Bellazzini:2015cgj} on the QCH in this channel.\footnote{We thank Dorival Gon\c{c}alves and Tao Han for alerting us to this discrepancy.} When $p>\mu$, the propagator and the vertices generate a non-integer power of minus one, $(-1)^{\nu}$, which is a multi-valued; the usual $i \epsilon$ prescription selects the correct branch . That is, for $p>\mu$,
\begin{eqnarray}
(\mu^2-p^2-i\epsilon)^{\nu}=(p^2-\mu^2)^{\nu}(-1-i \epsilon)^{\nu}&=& (p^2-\mu^2)^{\nu}e^{\log(-1-i\epsilon)^{\nu}}\nonumber\\
&=&(p^2-\mu^2)^{\nu}e^{+i \nu \pi}~.
\label{epsilon}
\end{eqnarray}
 This prescription correctly gives a positive spectral density for the propagator (\ref{densityfunction}). In the AdS picture, this prescription correctly gives outgoing waves in the IR after solving the equation in Euclidian space and Wick rotating back to Minkowski space-time.

 Furthermore, the same expression, (\ref{epsilon}), shows up in the $HZZ$ form factor.  Since Ward-Takahashi identities relate the form factor to the inverse propagator \cite{Mandelstam,Mandelstam2,Mandelstam3} we need to follow the  $i\epsilon$ prescription here as well. The consequence is that the sign of imaginary part of the Higgs diagram in $gg\to ZZ$ is flipped relative to leaving out the $i\epsilon$. In the off-shell region, SM matrix-elements of $gg\to H\to ZZ$  interfere destructively with the fermion box amplitude \cite{Campbell:2013una}. With an $i\epsilon$ in the form factor the QCH diagram interferes differently with the box diagram compared to a form factor without an $i\epsilon$. With an $i\epsilon$ the diagrams interfere destructively, as in the SM. Without an $i\epsilon$, the two diagrams interfere constructively. The large deviations observed previously arose from missing the $i\epsilon$ in the form factor.

\begin{figure}[H]
 \begin{subfigure}{0.5\textwidth}
 \centering
\includegraphics[width=1 \textwidth]{Madgraph_gg_H_ZZ.pdf}
\end{subfigure}
\begin{subfigure}{0.5\textwidth}
 \centering
\includegraphics[width=1 \textwidth]{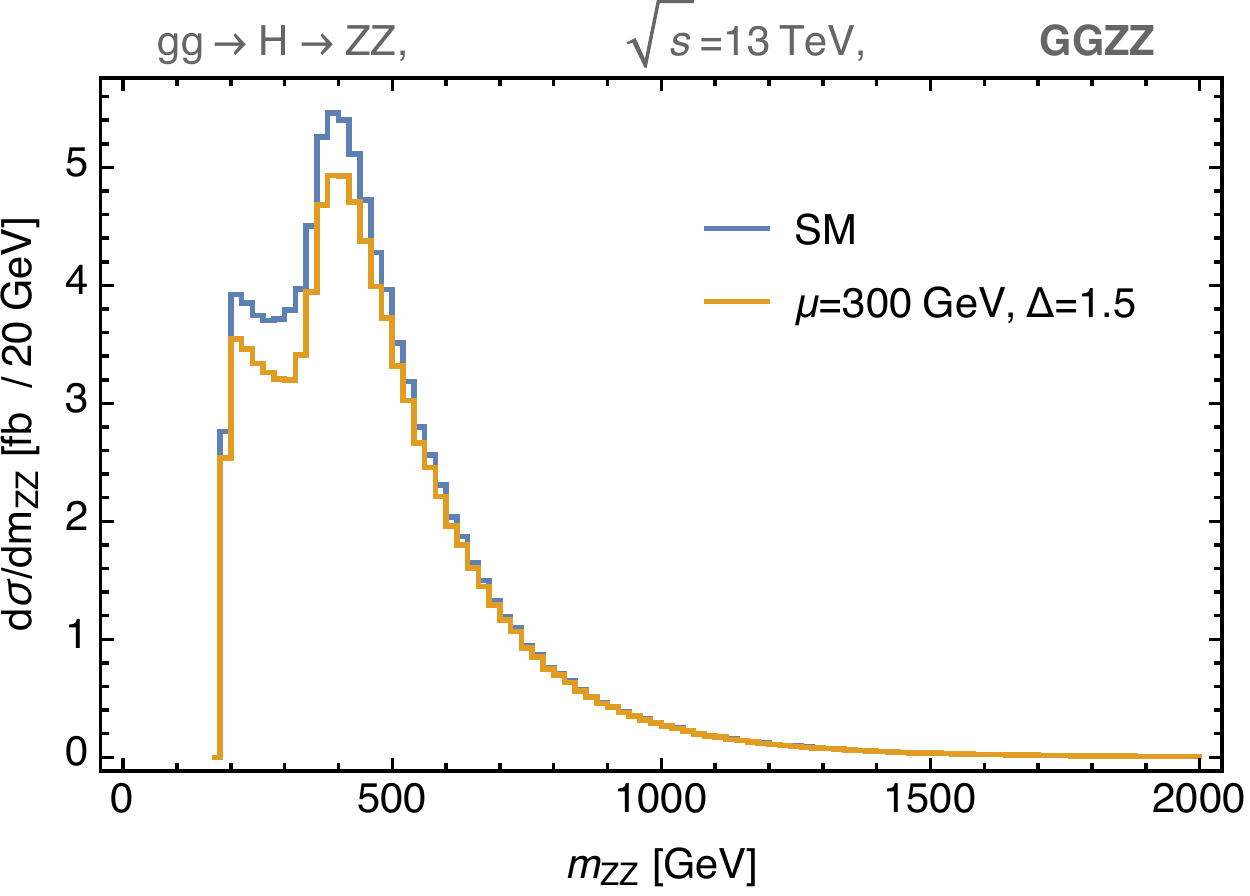}
\end{subfigure} 
\caption{$gg\to H \to ZZ$ versus the invariant mass of the two $Z$ bosons and with the $Z$ bosons on-shell.}
\label{fig:ggHZZ}
\end{figure}

\begin{figure}[H]
 \begin{subfigure}{0.5\textwidth}
 \centering
\includegraphics[width=1 \textwidth]{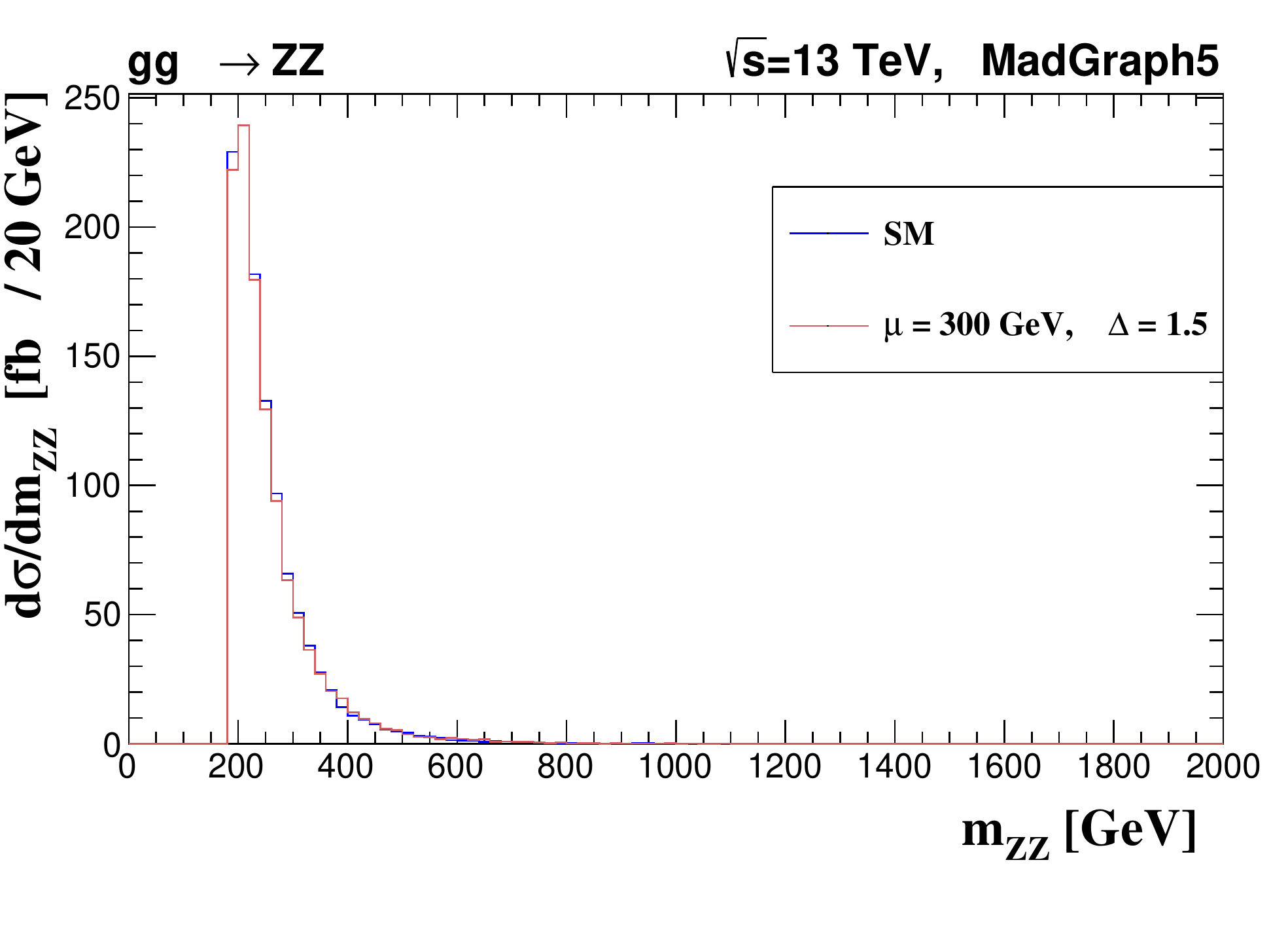}
\end{subfigure}
\begin{subfigure}{0.5\textwidth}
 \centering
\includegraphics[width=1 \textwidth]{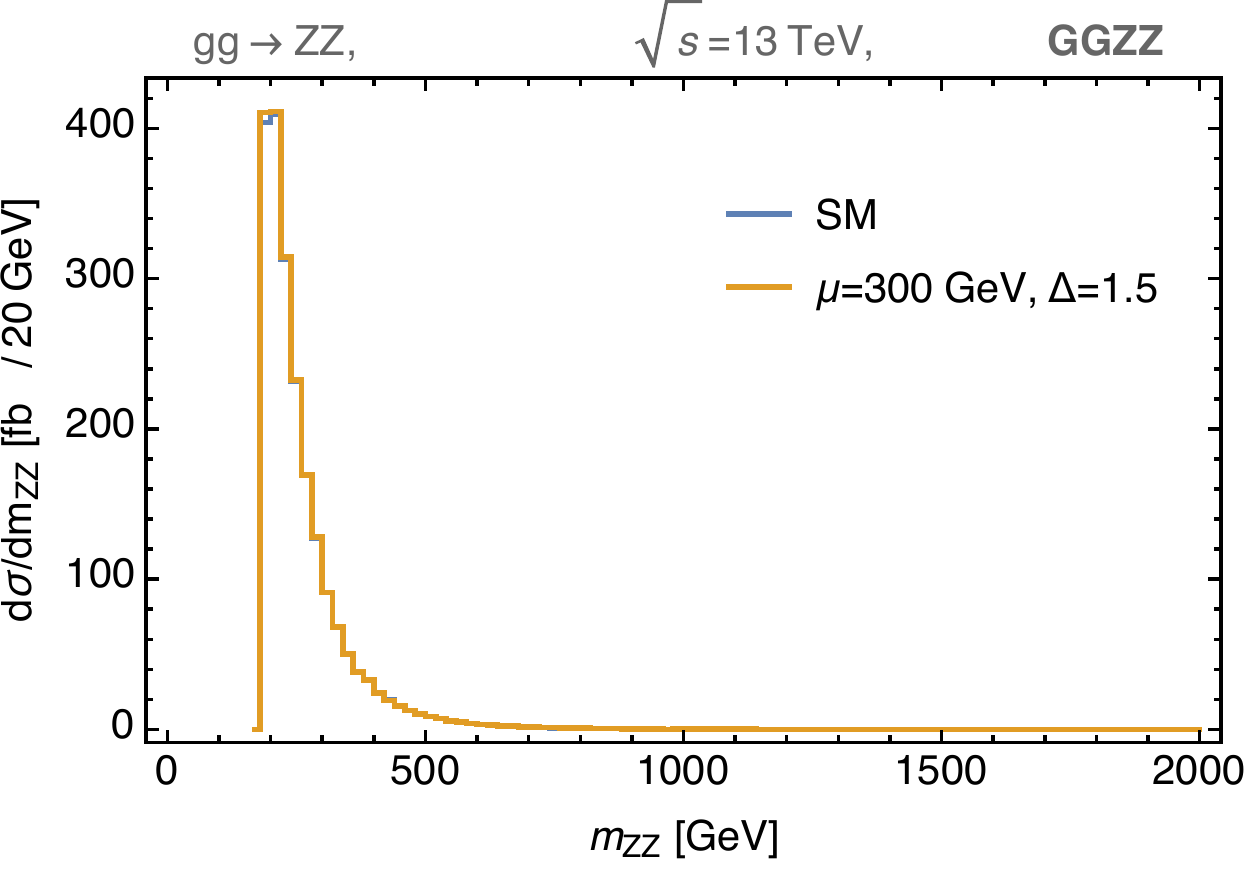}
\end{subfigure} 
\caption{$gg\to ZZ$. The sum of the box and Higgs diagrams with the $Z$ bosons on-shell.   }
\label{fig:ggZZtotal}
\end{figure}

\begin{figure}[H]
 \centering
\includegraphics[width=.7 \textwidth]{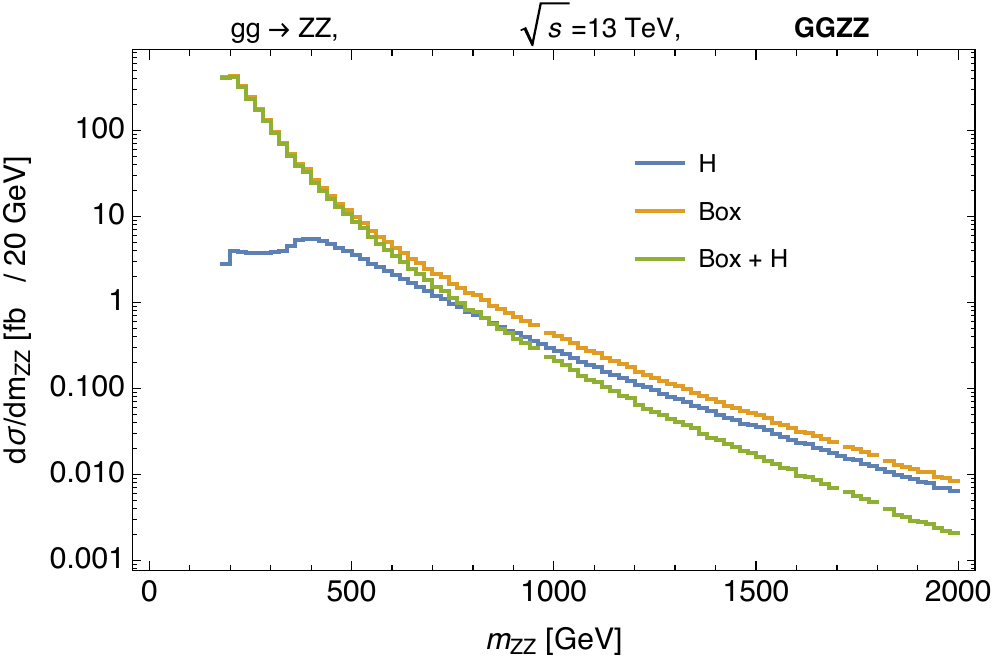}
\caption{$gg\to ZZ$. The interference of the Box and Higgs diagrams.   }
\label{fig:ggZZ}
\end{figure}

As a test of our implementation of MG we have compared the QCH in ZZ channel in MG and GGZZ\footnote{We thank Nigel Glover for providing his SM Fortran code.} \cite{Glover:1988rg,vanderBij:1988ac} program as shown in figures~\ref{fig:ggHZZ}, \ref{fig:ggZZtotal}, and \ref{fig:ggZZ}. Up to the overall normalization the results agree, and we can see that the implementation of QCH in MG works correctly. 


\subsection{$gg\to \gamma\gamma$}
\begin{figure}[ht!]
      \centering 
    \parbox{20mm}{\begin{fmffile}{ggaa}
        \begin{fmfgraph*}(100, 85)
         \fmftop{t1,t4}
          \fmfbottom{b1,b4}
          \fmf{curly}{t1,t2}
          \fmf{curly}{b1,b2}
          \fmf{fermion,tension=0.3}{t2,b2}
          \fmf{fermion}{b2,m1}
          \fmf{fermion}{m1,t2}
          \fmf{dbl_dashes}{m1,m2}
          \fmf{fermion}{m2,m3}
          \fmf{fermion,tension=0.3}{m3,m4}
          \fmf{fermion}{m4,m2}
          \fmf{photon}{m4,b4}
          \fmf{photon}{m3,t4}
        \end{fmfgraph*}\end{fmffile}}
    \caption{$gg\to H \to \gamma \gamma$}
    \label{fdiag:gghaa}
    \end{figure}
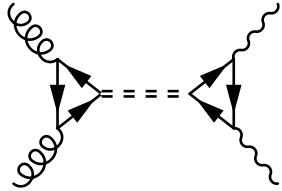

\begin{figure}[ht!]
\centering
\includegraphics[width=0.7\textwidth]{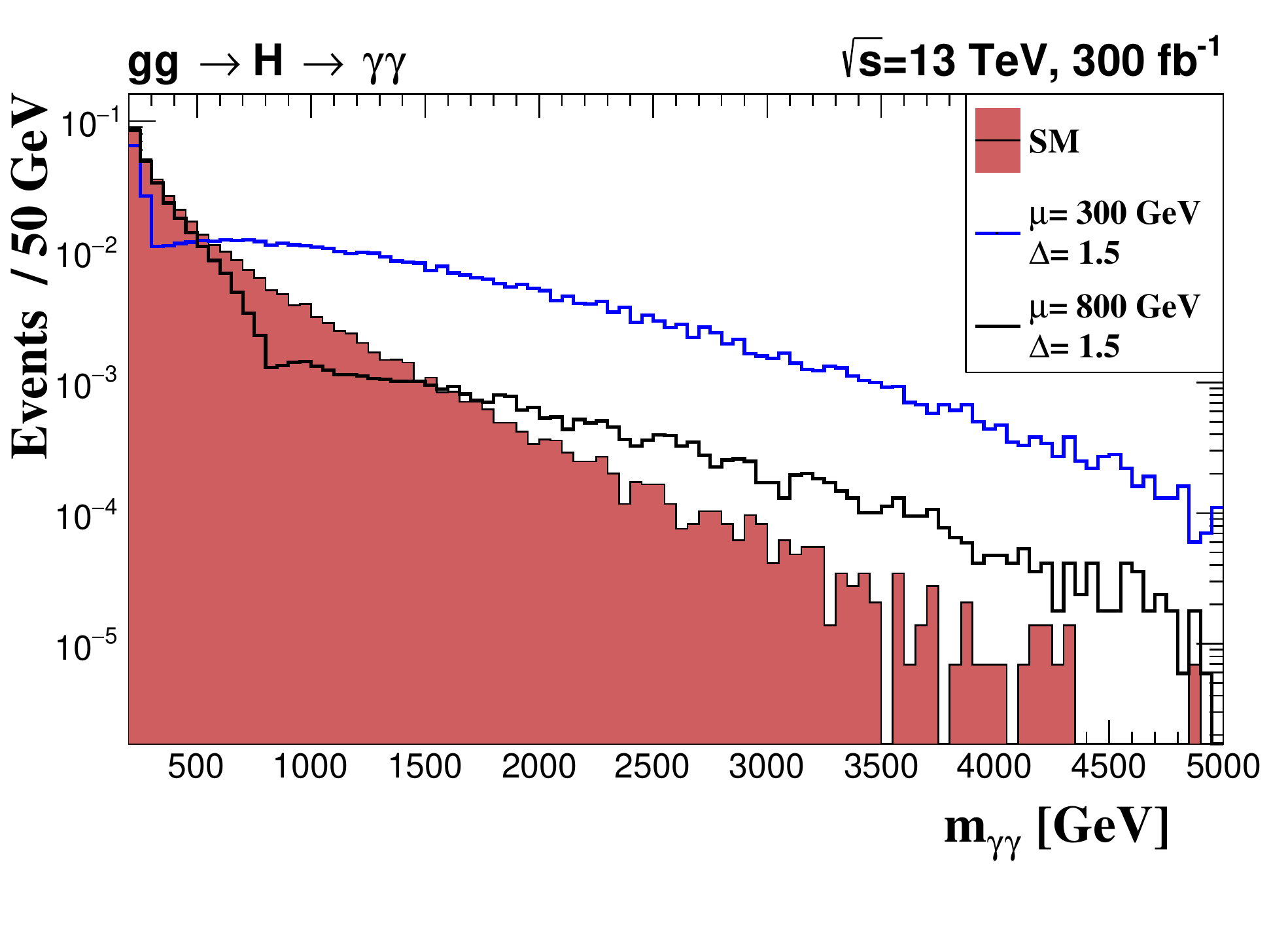}
  \caption{$gg\to \gamma\gamma$ at 13 TeV center of the mass energy with 300 fb$^{-1}$ integrated luminosity. }
  \label{fig:ggHaaMG}
 \end{figure}

\begin{figure}[ht!]
 \centering
\includegraphics[width=0.7\textwidth]{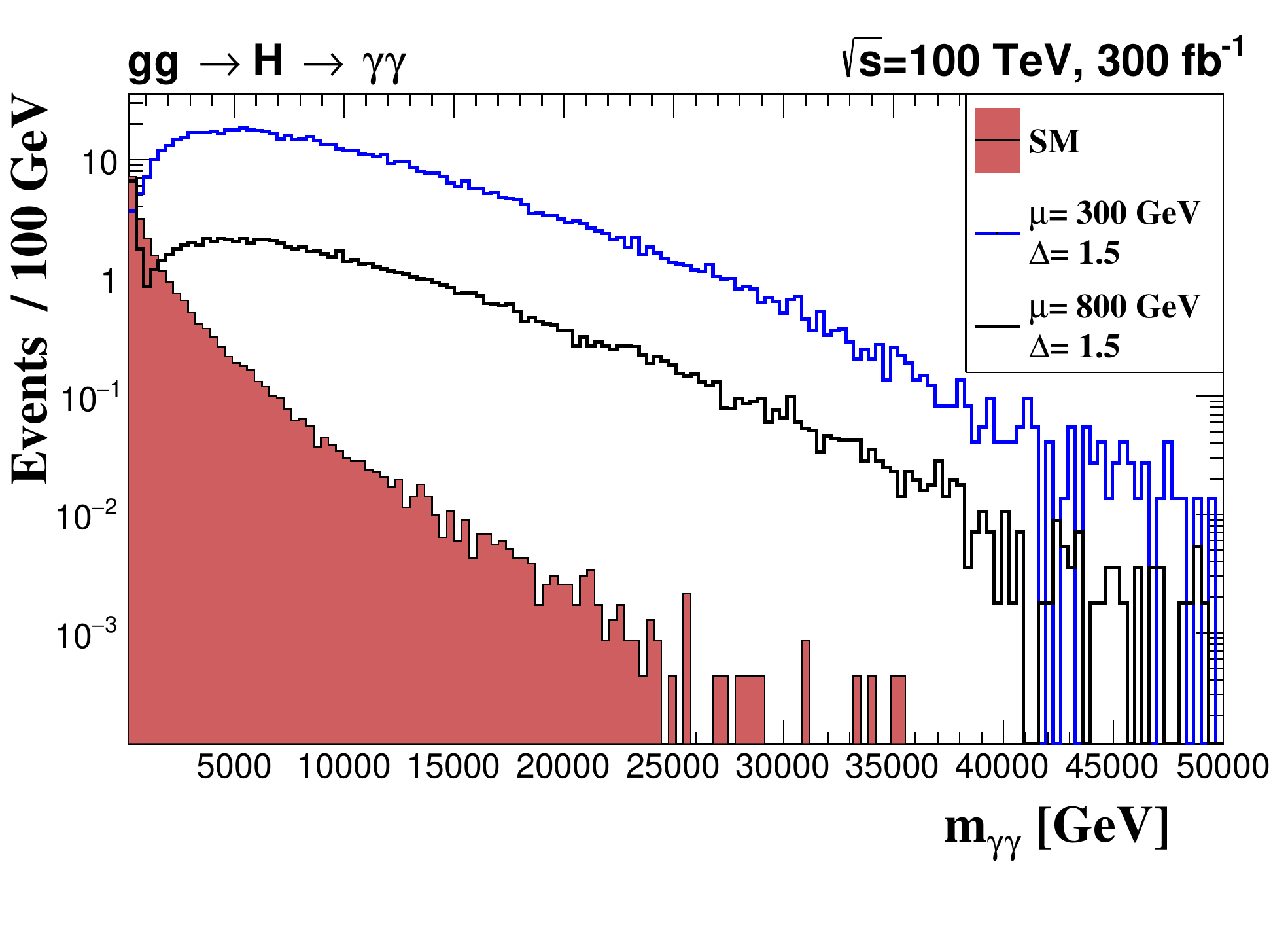}
\caption{$gg\to \gamma\gamma$ at 100 TeV center of the mass energy with 300 fb$^{-1}$ integrated luminosity.}
\label{fig:ggHaaMG100tev}
\end{figure}

The di-photon process (figure~\ref{fdiag:gghaa}) is another clean channel for studying Higgs and it has been studied in the off-shell region as well \cite{Aaboud:2017yyg}. Due to fact that there is no new form factor for $H\gamma\gamma$ interactions, the amplitude in this case probes the Higgs propagator, which in QCH models falls more slowly as $1/p^{2\nu}$. So the signal in this channel is characterized by an excess of events that grows with energy. Due to a complicated background analysis, we have only generated events using the QCH amplitude, as shown in figure~\ref{fig:ggHaaMG}, and postpone a more sophisticated analysis to future work. It is interesting to note that there are in fact a small excess of events has already observed in the high-mass region \cite{Aaboud:2017yyg}. However, even without any cuts, the cross section for the QCH with the parameter range we have explored is far too small to account for these events. A conclusive analysis can be done at a future hadron collider, where a very large number of events is predicted, as shown in figure~\ref{fig:ggHaaMG100tev}, in this channel.  
 

\newpage

\subsection{$gg\to HZ$}

Another interesting channel for testing QCH models is the production of a Higgs with a $Z$ or a $W$ boson \cite{Khachatryan:2016tnr}. This channel is the most sensitive production mode to study the decay of the Higgs boson to a pair of $b$ quarks, which is the dominant decay of the SM Higgs. Since these searches are optimized for the SM Higgs, the Higgs is not off-shell and so we are not probing continuum Higgs production. With the Higgs and $Z$ on-shell, this channel probes the $HZZ$  form factor. Taking the $Z$ off-shell we could also probe the continuum of longitudinal components of the gauge fields. We have assumed that the transverse components of gauge fields have a threshold far larger than the Higgs, but without this simplifying assumtion this channel would be sensitive to this continuum as well. 

For $HZ$ production the diagrams that contribute at leading order are drawn in figure~\ref{fdiag:gghz}.
 \begin{figure}[ht!]
      \centering
   \begin{eqnarray*}
\parbox{20mm}{\begin{fmffile}{ggzhbox}
        \begin{fmfgraph*}(100, 55)
         \fmftop{t1,t4}
          \fmfbottom{b1,b4}
          \fmf{curly}{t1,t2}
          \fmf{dashes}{t3,t4}
          \fmf{curly}{b1,b2}
          \fmf{photon}{b3,b4}
          \fmf{fermion}{t3,t2}
          \fmf{fermion,tension=0.3}{t2,b2}
         \fmf{fermion}{b2,b3}
          \fmf{fermion,tension=0.3}{b3,t3}
        \end{fmfgraph*}\end{fmffile}
    }
    \qquad\qquad +\quad\text{crossings}\quad+\qquad 
    \parbox{20mm}{\begin{fmffile}{ggzh}
        \begin{fmfgraph*}(100, 55)
         \fmftop{t1,t4}
          \fmfbottom{b1,b4}
          \fmf{curly}{t1,t2}
          \fmf{curly}{b1,b2}
          \fmf{fermion,tension=0.3}{t2,b2}
          \fmf{fermion}{b2,m1}
          \fmf{fermion}{m1,t2}
          \fmf{dbl_wiggly}{m1,m2}
          \fmf{photon}{m2,b4}
          \fmf{dashes}{m2,t4}
	 \fmfblob{0.15w}{m2}
        \end{fmfgraph*}\end{fmffile}}
        \end{eqnarray*}
    \caption{$gg\to hZ$}
    \label{fdiag:gghz}
    \end{figure}

\begin{figure}[ht!]
\centering
 \begin{subfigure}{0.49\textwidth}
\includegraphics[width=1 \textwidth]{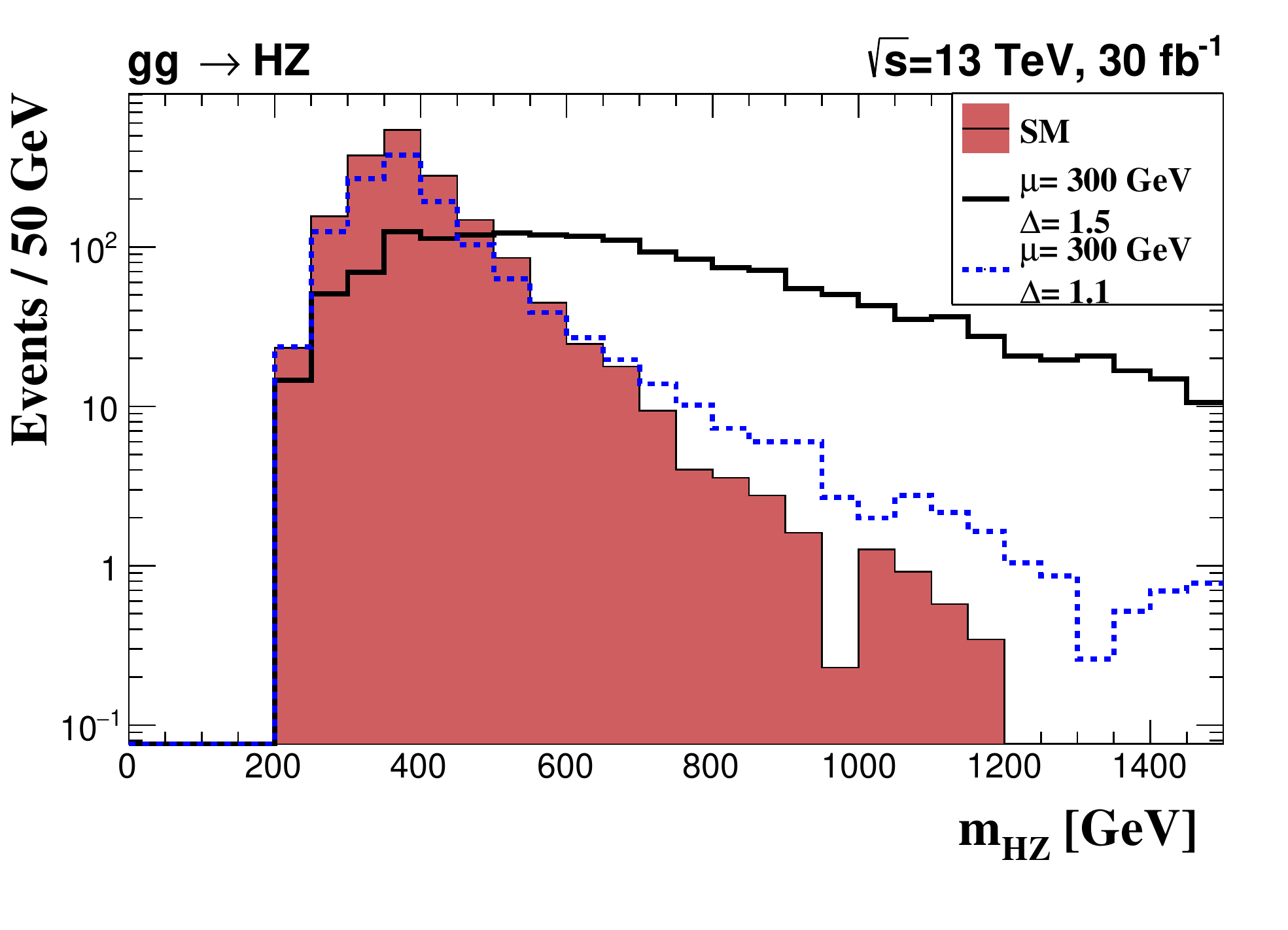}
     \end{subfigure}
   \begin{subfigure}{0.49\textwidth}
   \includegraphics[width=1 \textwidth]{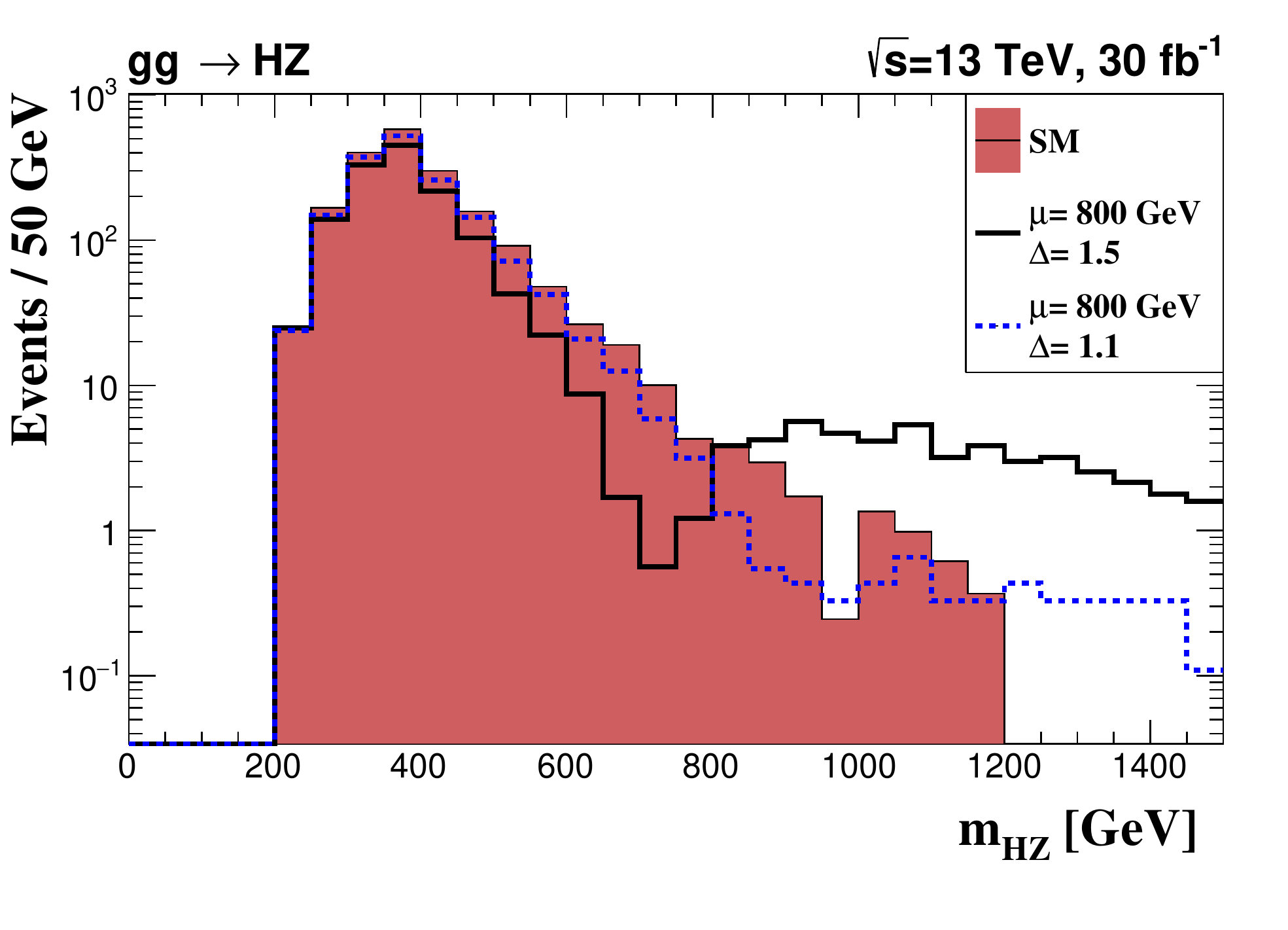}
\end{subfigure}
\caption{$gg\to HZ$, versus the invariant mass of Higgs and $Z$ boson, $m_{HZ}$.}
\label{fig:ggHZMHZ}
 \end{figure}

\begin{figure}[ht!]
 \begin{subfigure}{0.5\textwidth}
 \centering
\includegraphics[width=1 \textwidth]{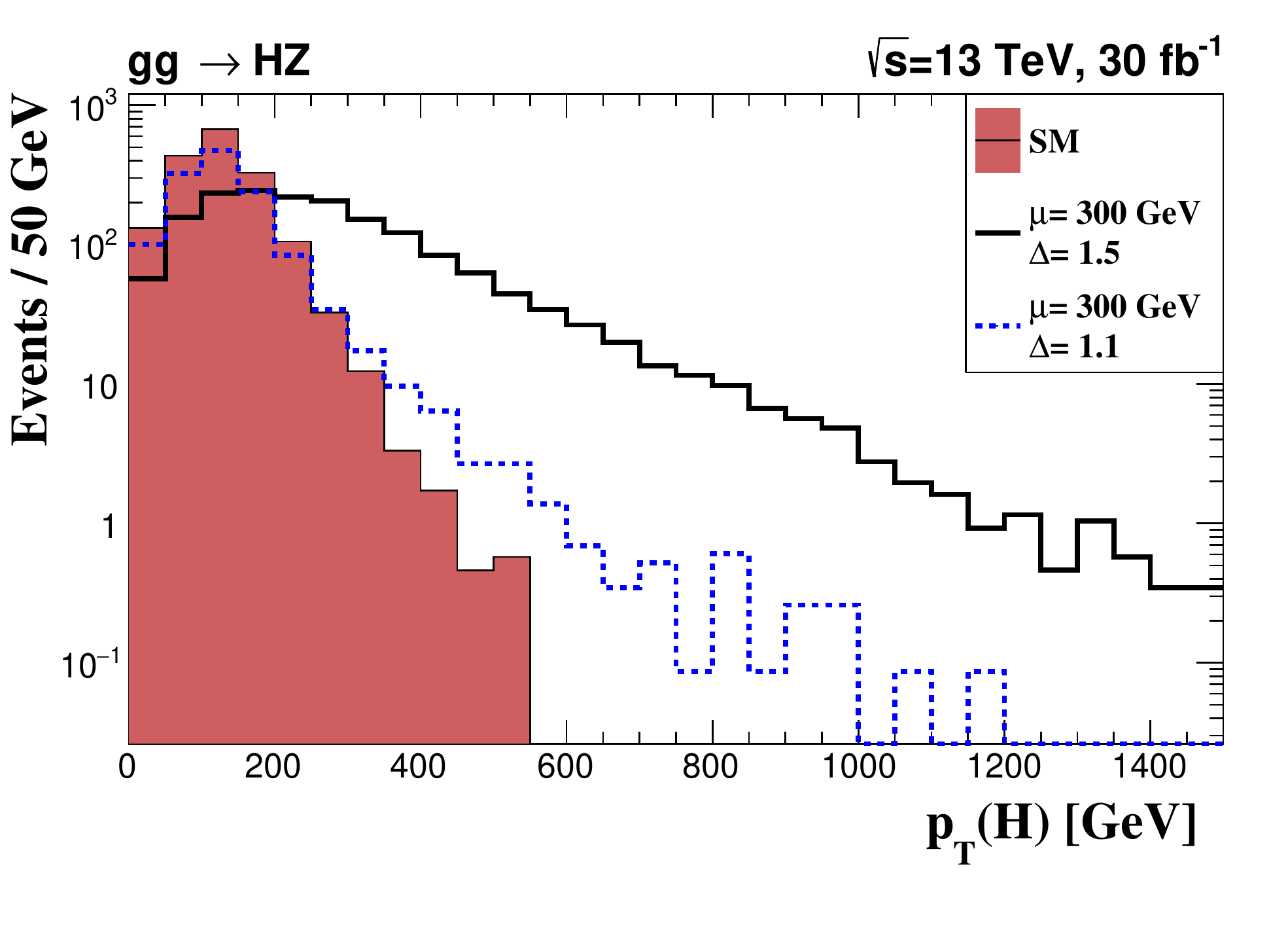}
\end{subfigure}
\begin{subfigure}{0.5\textwidth}
 \centering
\includegraphics[width=1 \textwidth]{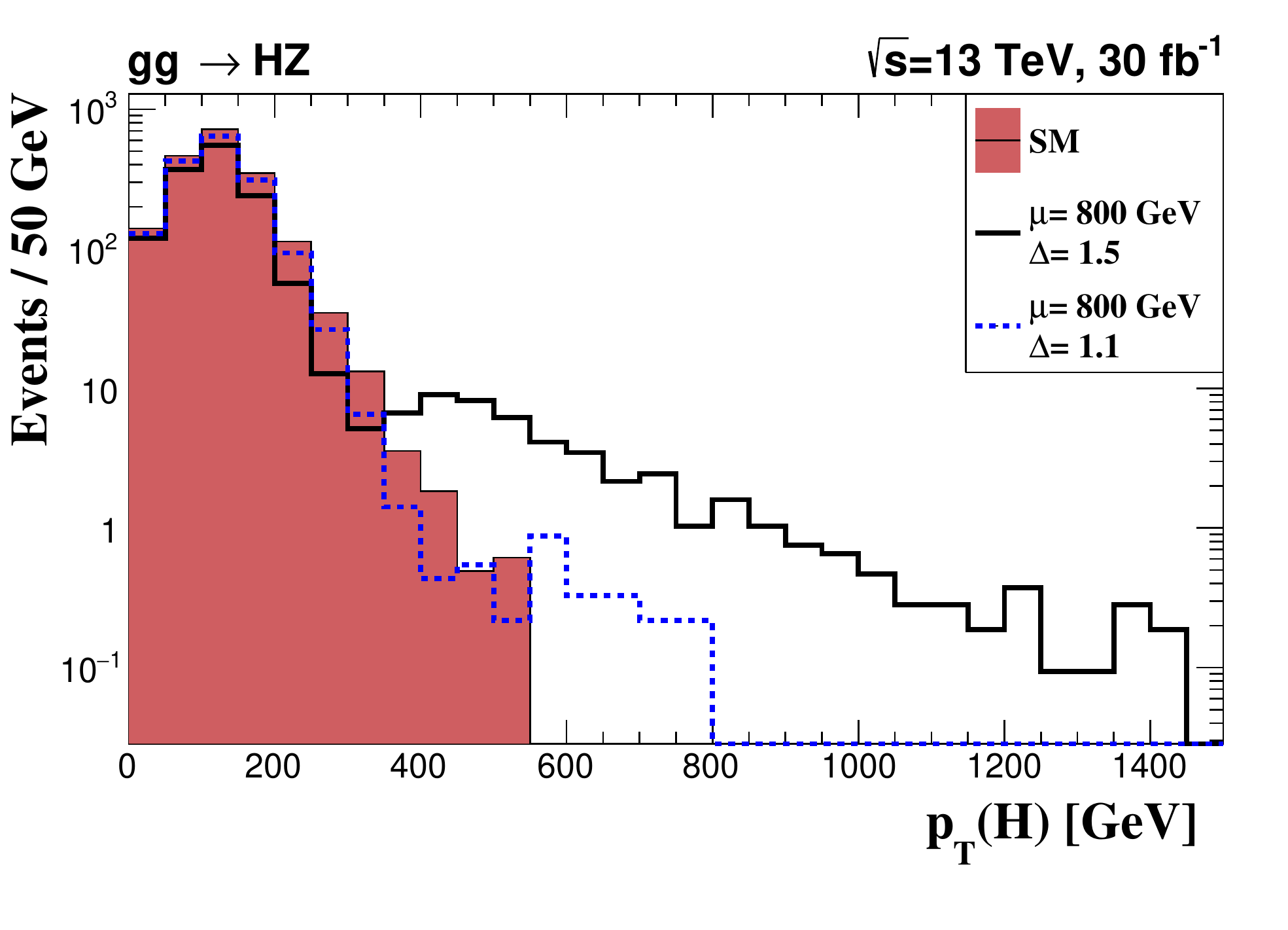}
\end{subfigure} 
\caption{$gg\to HZ$, versus the Higgs transverse momentum, $p_{T}(H)$}
\label{fig:ggHZPTH}
\end{figure}

\begin{figure}[ht!]
 \begin{subfigure}{0.5\textwidth}
 \centering
\includegraphics[width=1 \textwidth]{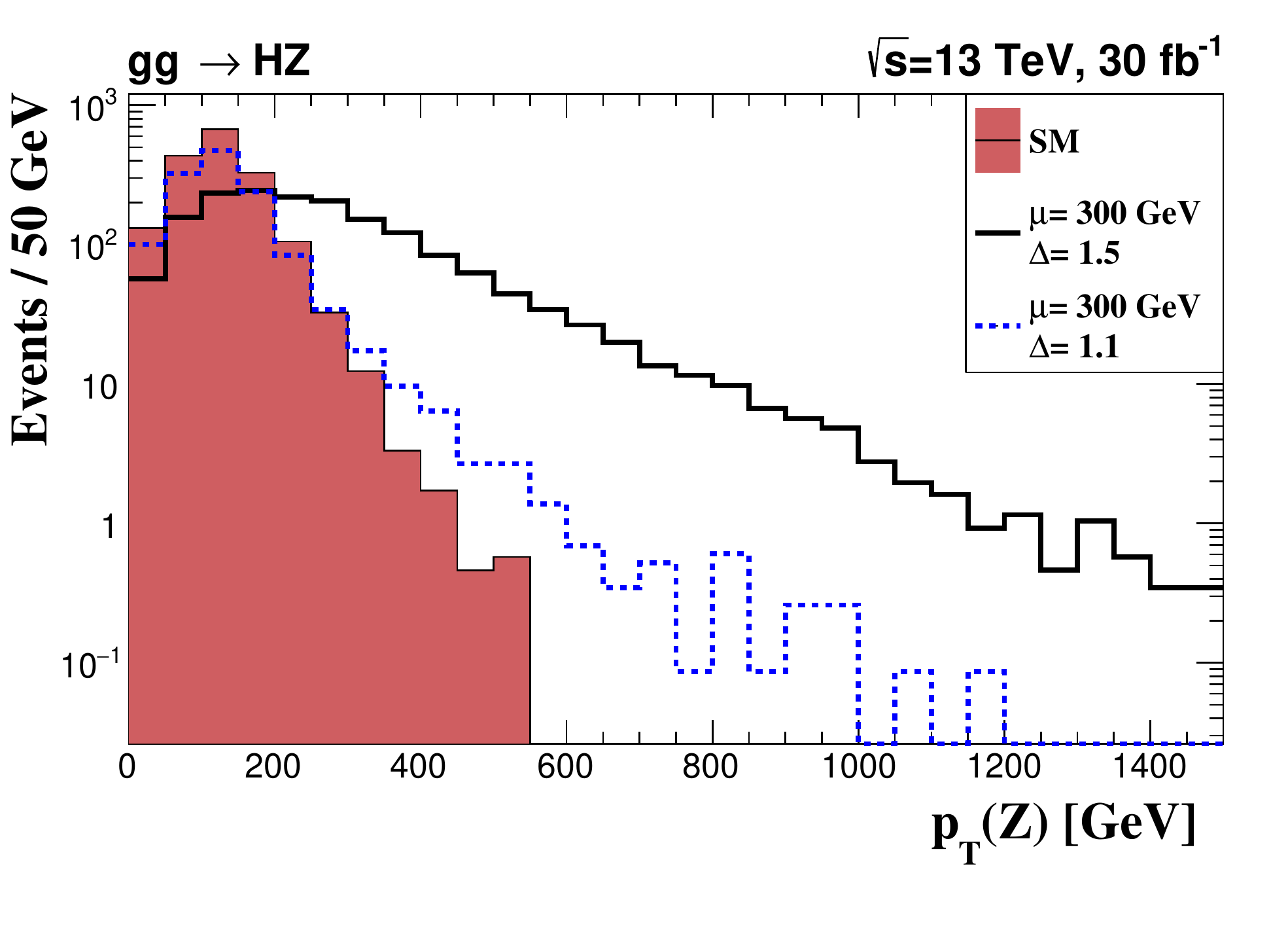}
\end{subfigure}
\begin{subfigure}{0.5\textwidth}
 \centering
\includegraphics[width=1 \textwidth]{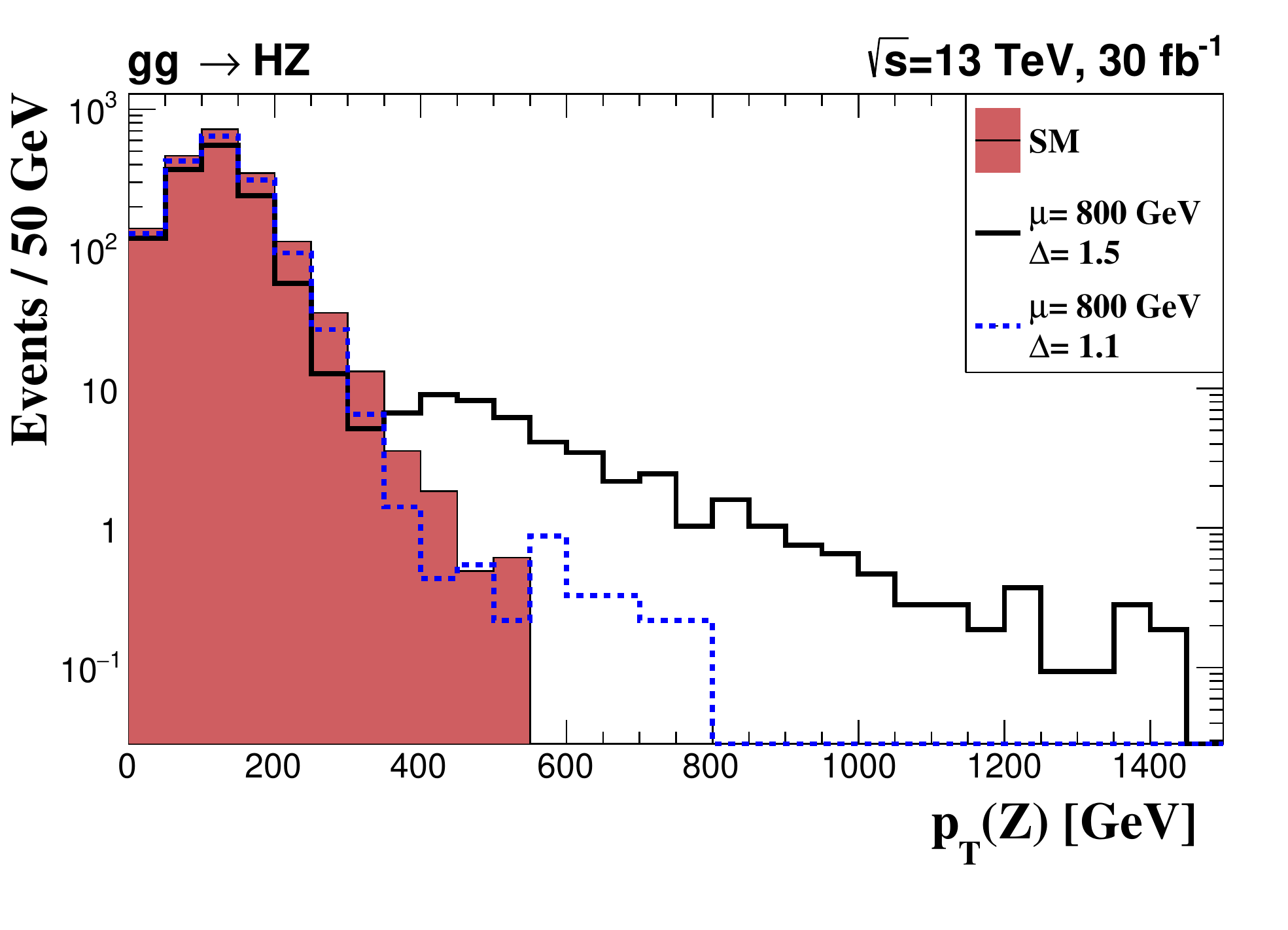}
\end{subfigure} 
\caption{$gg\to HZ$, versus the $Z$ boson $p_{T}(Z)$}
\label{fig:ggHZPTZ}
\end{figure}

There is a considerable deviation from the SM at small thresholds and large scaling dimensions as seen in Fig.\ref{fig:ggHZMHZ}. Aside from slight decrease in the cross-section below the threshold, the main signature is a considerable increase in the cross-section where SM cross-section becomes negligible. Even for the smaller luminosities than $300$ fb$^{-1}$, where the number of events in each bin might become small, especially after considering experimental cuts and efficiencies, the total number of observed event could be big enough for putting bounds on QCH. We will postpone a detailed comparison to experiment to future work. However, as seen in figure~\ref{fig:ggHZPTH} and figure~\ref{fig:ggHZPTZ}, we can see that imposing cuts on the transverse momenta will have a little effect on this behavior.

\section{Conclusions}

Although our approach is mainly phenomenological, we have presented two consistent 5D models for a QCH, in particular displaying their qualitatively similar features. We have further examined to what degree a QCH can reveal itself in current and the future experiments. In the $gg\to ZZ$ channel, the Higgs propagator and the $HZZ$ form-factor conspire to cancel each other's large deviations unless the threshold is at a special scale. The $gg\to \gamma\gamma$ channel seems to be promising for future colliders, however because of the complicated background we have not attempted a full analysis here. A promising channel for the LHC is $gg\to HZ$, where large deviations can be found in distributions over the invariant $HZ$ mass  and the transverse momenta of the $H$ and $Z$. Another channel that could be interesting to look at in the future is $gg \to W^{+}W^{-}$, since both the Higgs and the $Z$ boson thresholds can affect the observables in this channel.

\textbf{Acknowledgements} We thank Dorival Gon\c{c}alves, Tao Han, and Subir Sachdev and  for useful discussions. AS would like to also thank Rui Zheng and Nichol\'{a}s Neil, especially for helping with implementing the model in {\tt MadGraph}. This work was supported in part by DOE grant DE-SC0009999. 


\appendix
\section{The spectrum of scalar fields\\ In AdS/Broken-CFT}\label{ads/brokenCFT}
The metric of a 5D theory foliated with flat 4D space-times can be written as
\begin{equation}
ds^2=e^{-2A(z)}(\eta_{\mu\nu}dx^{\mu}dx^{\nu}-dz^2),
\end{equation}
where for an AdS background $A(z)=\log z$. Here we assume the $A(z)$ might deviate from AdS background. The action for a scalar field is
\begin{equation}
S\supset  \frac{1}{2}\int d^4xdz\sqrt{g}\Big[\partial_Mh\partial^Mh-M(z)^2h^2\Big],
\end{equation}
where we have assumed that the mass term might be a function of a coordinate dependent VEV. There is a nice trick \cite{RS} for figuring out the spectrum in such models. We substitute $h\to e^{3/2A}\bar{h}$ in \eqref{bulkhiggsequationofmotion}, and the result is a Schr\"{o}dinger like equation
\begin{align}
-\bar{h}''&+V\bar{h}=p^2\bar{h},\label{schrodinger}\\
V=&\frac{9}{4}A'^2-\frac{3}{2}A''+e^{-2A}M^2.\label{effPotential}
\end{align}
Depending on the background, there can be three scenarios:
\begin{itemize}
\item \textbf{Discretuum}. $V(z)\to \infty$ as $z\to \infty$. This is the case when there exists an IR brane, as in RS1 models \cite{RS}, or for certain soft-wall models in which the warp factor decays fast $A(z)\propto z^{\alpha},\alpha>1$. The singularity in the IR, a brane or a soft-wall, signals the breaking of the CFT in the IR. It is a hard breaking, the theory confines and we get the KK modes, the composites of the broken CFT. The mass splitting goes roughly as $\Delta m_{KK}=\frac{\pi}{z_{IR}}$, where $z_{IR}$ is the position of the brane or the soft-wall.
\item \textbf{Continuum without a mass gap}.  $V(z)\to 0$ as $z\to \infty$. This is the case with the AdS background, dual to an unbroken CFT, as in for example RS2 model \cite{Randall:1999vf}.
\item \textbf{Continuum with a mass gap}.  $V(z)\to\mu^2>0$ as $z\to \infty$. The dual CFT is softly broken such that the continuum is untouched but starts at a finite threshold, $\mu$. This behavior has been studies in  a variety of scenarios \cite{Falkowski:2008yr,Cacciapaglia:2008ns,Megias:2019vdb,Cai:2009ax} 
\end{itemize}

\section{Backreaction of the scalar field In the minimal AdS/QCH model}\label{app:backreaction}
Here we examine the back reaction of the field $\phi$ on the metric by finding the solution to the Einstein equations in the presence of the scalar field such that in the UV the metric asymptotically behaves as \eqref{backgroundmetric}, with $A(z)=\log z/R$, and the scalar field as \eqref{phi}.
The Einstein equations together with the equation of motion for the $\phi$ actually come down to two independent equations
\begin{align}
A''(z)-\frac{1}{4M_5^3}\phi'(z)^2-\frac{1}{12M_5^3}e^{-2A(z)}V(\phi)=&0\\
\phi''(z)-3A'(z)\phi'(z)-\frac{1}{2}e^{-2A(z)}\partial_{\phi}V(\phi)=&0,
\end{align}
where $M_5$ is the 5D Plank mass. To reproduce the AdS background near the UV and the correct scaling of the  the potential is
\begin{equation}
V=-\frac{12M_5^3}{R^2}-\frac{4}{R^2}\phi^2.
\end{equation}

 Unfortunately, there is no known analytic solution to the equations above. Solving the equations numerically we find that there is a singularity\footnote{Close to the singularity the derivative of the warped factor is 
 \begin{equation}
 A'(z)=\frac{1}{3(z_s-z)}.
 \end{equation}
} at
\begin{equation}
z_s\approx \frac{1.335(RM_5)^{3/4}}{\mu}
\end{equation}
The appearance of this singularity has the dual interpretation that there is a hard breaking of the symmetry; the corresponding KK modes are dual to composites of the CFT. Of course whether the singularity is actually a hard breaking or just a coordinate artifact depends on the boundary conditions.  Assuming a hard breaking the spectrum is a tower of KK modes starting at $\mu$ with mass splitting given by $\Delta m_{KK}\approx \pi/z_s$. To mimic a continuum, one can adjust the AdS curvature radius, $R$, such that  $R\gg 1/M_5$ and $\Delta m_{KK} \leq \Gamma_H$, the Higgs width. This is similar to a Pad\'{e} approximation of the continuum \cite{PerezVictoria:2008pd}. 

It is to be expected that the spontaneous breaking of scale invariance also comes with a hard breaking, the renormalization flow generally leaves the vicinity of the fixed point at some point. The fact that there is no such singularity in the other two models analyzed above can be viewed as a tuning of parameters such that this singularity is at infinity.

 \section{QCH in {\tt MadGraph}}\label{app:qchmadgraph}
In order to extend {\tt MadGraph}, which calculates SM amplitudes to the next-to-leading order, to QCH models we need to change the propagators and add momentum dependent vertices. We do this by extending the \emph{loop\_sm} model which exist in MG directory. We add the new propagators in a file \emph{propagator.py}.\footnote{This file and \emph{formfactor.py} can be found in other BSM models in MG but not \emph{loop\_sm}.} For example, the scalar field the propagator reads
 \begin{verbatim}
S = Propagator(name = "S",
      numerator = "(2-dim)/(cut**2-Mass(id)*Mass(id))**(dim-1)",
      denominator = "-pow(cut**2 - P('mu', id) * P('mu', id) 
      - complex(0,1) * eps,2.0-dim)+pow(cut**2 - Mass(id) * 
       Mass(id),2.0-dim) + complex(0,1) * Mass(id) * Width(id)
       *(2-dim)/(cut**2-Mass(id)*Mass(id))**(dim-1)"
        )
  \end{verbatim}
 For convenience we defined new parameters \emph{cut} ($=\mu$) and \emph{dim} ($=\Delta=2-\nu$) in the files \emph{parameters.py} and \emph{restrict\_default.dat}, so that we can easily change these from within MG. For example, after launching the program, these parameters will show up in the \emph{parameter} card and can be changed by the command, e.g.
 \begin{verbatim}
 set dim 1.8
 \end{verbatim}

  We then declare the new propagators for the particles in \emph{particle.py}. As an example, for the Higgs we have
  \begin{verbatim}
  import propagators as Prop
 H = Particle(pdg_code = 25,
             name = 'H',
             antiname = 'H',
             spin = 1,
             color = 1,
             mass = Param.MH,
             width = Param.WH,
             texname = '\\phi',
             antitexname = '\\phi',
             charge = 0,
             LeptonNumber = 0,
             GhostNumber = 0,
             propagator = Prop.S)
  
  \end{verbatim}

In a similar fashion the form factors are defined in \emph{formfactor.py} and passed through the vertices through the files \emph{lorentz.py} and \emph{vertices.py}.
 
 Since the model \emph{loop\_sm} does not accept \emph{propagator.py} and \emph{formfactor.py} by default, we need to change two more files. In \emph{ \_init\_.py } we add the lines
 \begin{verbatim}
all_propagators = propagators.all_propagators
all_form_factors = form_factors.all_form_factors
 \end{verbatim}

  and in \emph{object\_library.py} we add \emph{propagator} as another argument to the \emph{particle} class, and define \emph{Propagator} and \emph{FormFactor} classes.


\end{document}